\documentclass[aip,pre,11pt]{revtex4-1}
\usepackage[a4paper, total={7in, 10in}]{geometry}
\usepackage{subcaption}
\usepackage{graphicx}
\usepackage{dcolumn}
\usepackage{bm}
\usepackage{xcolor, soul}
\usepackage{amsfonts}
\usepackage{booktabs}
\usepackage{siunitx}
\usepackage{ragged2e}
\usepackage[colorlinks,citecolor=red,urlcolor=blue,bookmarks=false,hypertexnames=true]{hyperref} 
\usepackage{float}
\usepackage{float}
\usepackage{xcolor}
\usepackage{array}
\usepackage{setspace}
\usepackage{lipsum}
\usepackage[font=small]{caption}
\usepackage{amssymb}
\usepackage{bm}

\usepackage[colorlinks,citecolor=red,urlcolor=blue,bookmarks=false,hypertexnames=true]{hyperref} 

\newcommand{\HM}[1]{{\color{black}{#1}}}

\newcolumntype{P}[1]{>{\centering\arraybackslash}p{#1}}
\newcolumntype{M}[1]{>{\centering\arraybackslash}m{#1}}
\usepackage{setspace}
\usepackage{amsmath}
\usepackage{parskip}

\begin{document}




\title{Magnetophoresis of Weakly Magnetic Nanoparticle Suspension Around a Wire}

\author{Mohd Bilal Khan}
\affiliation{Department of Chemical and Biomedical Engineering, FAMU-FSU College of Engineering, Tallahassee, FL, 32310, USA}
\affiliation{Center for Rare Earths, Critical Minerals, and Industrial Byproducts, National High Magnetic Field Laboratory, Tallahassee, FL 32310, USA}

\author{Peter Rassolov}
\affiliation{Department of Chemical and Biomedical Engineering, FAMU-FSU College of Engineering, Tallahassee, FL, 32310, USA}
\affiliation{Center for Rare Earths, Critical Minerals, and Industrial Byproducts, National High Magnetic Field Laboratory, Tallahassee, FL 32310, USA}
\author{Jamel Ali}
\affiliation{Department of Chemical and Biomedical Engineering, FAMU-FSU College of Engineering, Tallahassee, FL, 32310, USA}
\affiliation{Center for Rare Earths, Critical Minerals, and Industrial Byproducts, National High Magnetic Field Laboratory, Tallahassee, FL 32310, USA}
\author{Theo Siegrist}
\affiliation{Department of Chemical and Biomedical Engineering, FAMU-FSU College of Engineering, Tallahassee, FL, 32310, USA}
\affiliation{Center for Rare Earths, Critical Minerals, and Industrial Byproducts, National High Magnetic Field Laboratory, Tallahassee, FL 32310, USA}
\author{Munir Humayun}
\affiliation{Center for Rare Earths, Critical Minerals, and Industrial Byproducts, National High Magnetic Field Laboratory, Tallahassee, FL 32310, USA}
\affiliation{Department of Earth, Ocean and Atmospheric Science, Florida State University, Tallahassee, FL 32304, USA.}

\author{Hadi Mohammadigoushki}
\thanks{Corresponding author}\email{hadi.moham@eng.famu.fsu.edu}
\affiliation{Department of Chemical and Biomedical Engineering, FAMU-FSU College of Engineering, Tallahassee, FL, 32310, USA}
\affiliation{Center for Rare Earths, Critical Minerals, and Industrial Byproducts, National High Magnetic Field Laboratory, Tallahassee, FL 32310, USA}
\date{\today}

\begin{abstract}

We present a combined experimental and numerical investigation into the magnetophoresis behavior of weakly magnetic nanoparticle suspensions in the vicinity of a wire under a non-uniform magnetic field and negligible inertia. The experiments were conducted within a closed rectangular cuvette, with a wire positioned between the poles of an electromagnet. Two types of nanoparticles—paramagnetic manganese oxide and diamagnetic bismuth oxide—were studied across a broad range of concentrations (10–100 mg/L), magnetic field strengths (0.25–1 T), and wire diameters (0.8–3.17 mm). Our experimental findings reveal that, upon the application of a magnetic field, paramagnetic nanoparticles experience a strong, attractive force toward the wire's periphery. This force generates vortices and secondary flows around the wire, depleting particles from the bulk of the cuvette and concentrating them near the wire surface. The magnetophoresis dynamics of paramagnetic nanoparticles are shown to scale with their initial concentration, wire diameter, and the strength of the external magnetic field. In contrast, diamagnetic nanoparticles exhibit markedly different behavior, with their magnetophoresis dynamics showing minimal dependence on initial concentration and magnetic field strength, while being inversely proportional to the wire diameter. Multiphysics numerical simulations complement the experimental observations, revealing the formation of field-induced particle clusters in weakly paramagnetic nanoparticles, which enhance magnetophoresis. Additionally, the critical magnetic field threshold for the onset of cluster formation is found to be lower than those predicted by theoretical models for clustering in uniform magnetic fields. Under specific conditions, including high magnetic field strengths and elevated nanoparticle concentrations, diamagnetic nanoparticles appear to undergo field-induced clustering, suggesting a previously unreported aspect of their magnetophoretic behavior.

\end{abstract}
\maketitle




\section{Introduction}

{Transport and separation processes are fundamental to a wide range of applications, from water purification\cite{li2012application}, resource recovery\cite{tseng2023application} to environmental protection\cite{yang2017application}, biotechnology~\cite{schwaminger2019magnetic}, medical diagnostics~\cite{leong2016working} amongst others. Today, many effective and well-established separation approaches exist, including mechanical, thermal, chemical, electrical, and/or magnetic, each with its own limitations based on the physical properties involved in its working principle~\cite{cui2003mechanical,sattler2008thermal,manouchehri2000review,svoboda2004magnetic}. Among various separation techniques, magnetic separation (or magnetophoresis) offers several advantages, including being energetically efficient, environmentally benign, and having high selectivity with minimal wear and tear\cite{svoboda2003recent,kemsheadl1985magnetic,nithya2021magnetic,munaz2018recent}.  } Magnetophoresis relies on the principles of magnetism to selectively separate materials based on their magnetic properties. {Under the influence of an applied magnetic field, a particle with no net charge, experiences a magnetic force that is directly proportional to magnetic susceptibility of the particle and the magnetic field gradients~\cite{rassolov2024magnetophoresis,butcher2022magnetic,boyer1988force,xue2014template}. Magnetic materials are categorized into paramagnetic and diamagnetic types based on their magnetic susceptibility ($\chi$), with paramagnetic materials exhibiting positive susceptibility ($\chi>$ 0) and diamagnetic materials showing negative susceptibility ($\chi <$ 0)\cite{svoboda2004magnetic}.\par


Early research by several groups underscored the importance of magnetic field gradients in separating magnetic and non-magnetic materials, laying the groundwork for High-Gradient Magnetic Separation (HGMS)~\cite{svoboda2004magnetic,oberteuffer1973high,hayashi2010development,kim2013effects,podoynitsyn2016high}. {High-gradient magnetic separators (HGMS) enhance the magnetic field gradient by incorporating a mesh of randomly oriented ferromagnetic wires or parallel plates~\cite{huang2015study,zeng2019selective}. This mesh serves as the magnetic matrix, generating highly localized and intense magnetic fields when placed inside a uniform magnetic field. HGMS has been used to capture strongly magnetic (superparamagnetic) micro and nano-sized particles based on their magnetic susceptibility~\cite{ngomsik2005magnetic,hwang1984application}. Studies on high-gradient magnetic separation have shown that the separation efficiency of micro and nano-particles in such systems depends on a coupling between forces involved (e.g., inertia, viscous, magnetic, diffusion, and/or surface forces), and the mesh type and size \cite{munaz2018recent,ge2017magnetic}. More recently, Benhal et al.~\cite{prateek25} investigated the separation of transition metal ions using an HGMS design. Their findings demonstrated that paramagnetic metal ions can be effectively separated from diamagnetic ions within an HGMS system~\cite{prateek25}. Despite advances in high-gradient magnetic separation (HGMS) for micro- and nano-particles with strong magnetic properties, several challenges remain that hinder a systematic understanding of the transport processes involved. A significant issue is the heterogeneity of the ferromagnetic matrix, which consists of randomly oriented wires. This randomness can lead to spatio-temporal variations in magnetic capture efficiency, complicating accurate modeling of the transport and fluid flow under an external magnetic field. Additionally, mesh-based systems often encounter clogging issues when processing fine particles, leading to reduced efficiency and operational challenges~\cite{svoboda2004magnetic,stephens2012analytical,ge2017magnetic}.}\par




{To address the complexities of HGMS and gain a deeper understanding of the underlying mechanisms involved in this process, several studies have focused on the hydrodynamic interactions between magnetic microparticles and a single wire, specifically examining the flow of particles past a magnetized wire~\cite{cowen1977single,friedlaender1978particle,mcneese1979viscosity,wankat1984removal,krafcik2019high,bilgili2022modeling,nameni2022separation}. This approach simplifies the system by isolating the behavior of magnetic particles in proximity to a single ferromagnetic element.} On the theoretical side, the first theoretical study was conducted by Watson in the early 70s~\cite{watson1975theory}. Therein, he derived an expression for the particle trajectory around a single wire under the assumption of steady-state inviscid flow (where inertia is significant).
Other theoretical studies examined the impact of microparticle accumulation on a single wire in high-gradient magnetic fields, and showed that as the magnetic field strength increases, the capture zone expands but eventually stabilizes with time~\cite{chen2012dynamic,choomphon2017simulation}. Experimental studies demonstrated that when the magnetic field is applied to the single ferromagnetic wire, micro-particle collection efficiency improves significantly, leading to more particle buildup on the wire surface~\cite{svoboda1988single}. Similarly, Friedlaender et al.~\cite{friedlaender1982study} showed that the magnetic field gradient creates a localized force of attraction around the wire, capturing the particles. Takayasu et al.~\cite{takayasu1983collection} studied the behavior of highly magnetized micro-particles under a high magnetic field gradient. They found that particle transport is influenced by both the magnetic properties and particle size. Smaller particles show less separation, and compact particle buildup occurs only when the fluid velocity is similar to the magnetic particle velocity. Wankat et al.~\cite{wankat1984removal} investigated the partial removal of paramagnetic micro-particles by reducing the magnetic field after initial buildup. The results revealed that particle buildup on the wire was saturated when the magnetic field was strong. \par

The above studies on single wire-suspended particle magnetic interactions have primarily focused on a flow of strongly magnetic (superparamagnetic) or micron-sized $\mathcal{O}({10-100}{\mu m})$ particles past a wire. In these studies, multiple forces—such as inertia, hydrodynamics (or bulk flow), and magnetic forces can concurrently influence particle transport. To the best of our knowledge, magnetophoresis of weakly paramagnetic and/or diamagnetic nano-particles around a magnetized wire has not been investigated experimentally or through numerical simulations. In the absence of inertia and hydrodynamic forces (no bulk flow), magnetic forces might be insufficient to overcome diffusive forces for weakly magnetic nano-particles, which could result in a behavior markedly different from that of larger or strongly magnetic particles~\cite{rassolov2024magnetophoresis,rassolov2025}. The primary aim of this study is to systematically investigate the magnetophoresis of weakly paramagnetic and diamagnetic nanoparticles in high-gradient magnetic fields, focusing exclusively on the competition between magnetic forces and diffusivity. To achieve this, both experiments and numerical simulations are conducted in a closed cuvette containing a nano-particle suspension and a ferromagnetic wire over a broad range of parameters including wire size, particle concentration, and external magnetic field strength. The spatio-temporal evolution of particle concentration is measured and compared with detailed multiphysics numerical simulations developed in this study. This direct comparison between experiments and simulations provides valuable insights into the underlying physics of magnetophoresis for weakly magnetic nanoparticle suspensions in non-uniform magnetic fields.

\section{\label{Methodology1}Experiments}

\subsection{{\label{Method}Materials}}

Paramagnetic manganese oxide particles with an average radius, $R_{p}$ = 50 nm, and magnetic susceptibility, $\chi$ = $14100 \times 10^{-6}$ cm$^{3}/$mole~\cite{lide2004crc} and diamagnetic bismuth oxide particles with an average radius, $R_{p}$ = 40 nm, and magnetic susceptibility of $\chi$ = $-83 \times 10^{-6}$ cm$^{3}/$mole~\cite{lide2004crc} are considered. {The particle density of the manganese oxide, $\rho$ = 5.03 g/cm$^{3}$ and while that of bismuth oxide, $\rho$ = 8.93 g/cm$^{3}$}. These nanoparticles were obtained from Sigma-Aldrich and used as received. Polyethylene Glycol (M = 6000 {[g/mole]}) is also obtained from Sigma Aldrich and used as received. The nanoparticle suspensions are prepared using deionized water. Due to its mechanical properties and corrosion resistance ability, stainless steel wire (430 grade, provided by McMaster-Carr) is used for the experiments. The radius of the wire varied from 0.8-3.17 mm within a standard size cuvette made up of polystyrene (12.5 mm $\times$ 12.5 mm $\times$ 45 mm).}

\subsection{{\label{Method}Solution preparation}}

To improve particle stability and prevent aggregation in solution, we functionalize the surface of nanoparticles with polyethylene glycol (PEG), a widely established method\cite{shrestha2020nanoparticle}. First, PEG is added at a concentration ten times that of the nanoparticles. The solution was allowed to sit for an hour, ensuring a complete dissolution of PEG in de-ionized water. Afterward, the solution was uniformly mixed using a vortex mixer. To further minimize the risk of particle aggregation, the solution was sonicated for 30 minutes in a bath sonicator, helping to break up any potential aggregates and maintain nanoparticle dispersion.  


\begin{figure}[hthp]
 \centering
 \includegraphics[trim=0cm 2cm 9cm 0cm,clip,width=14cm]{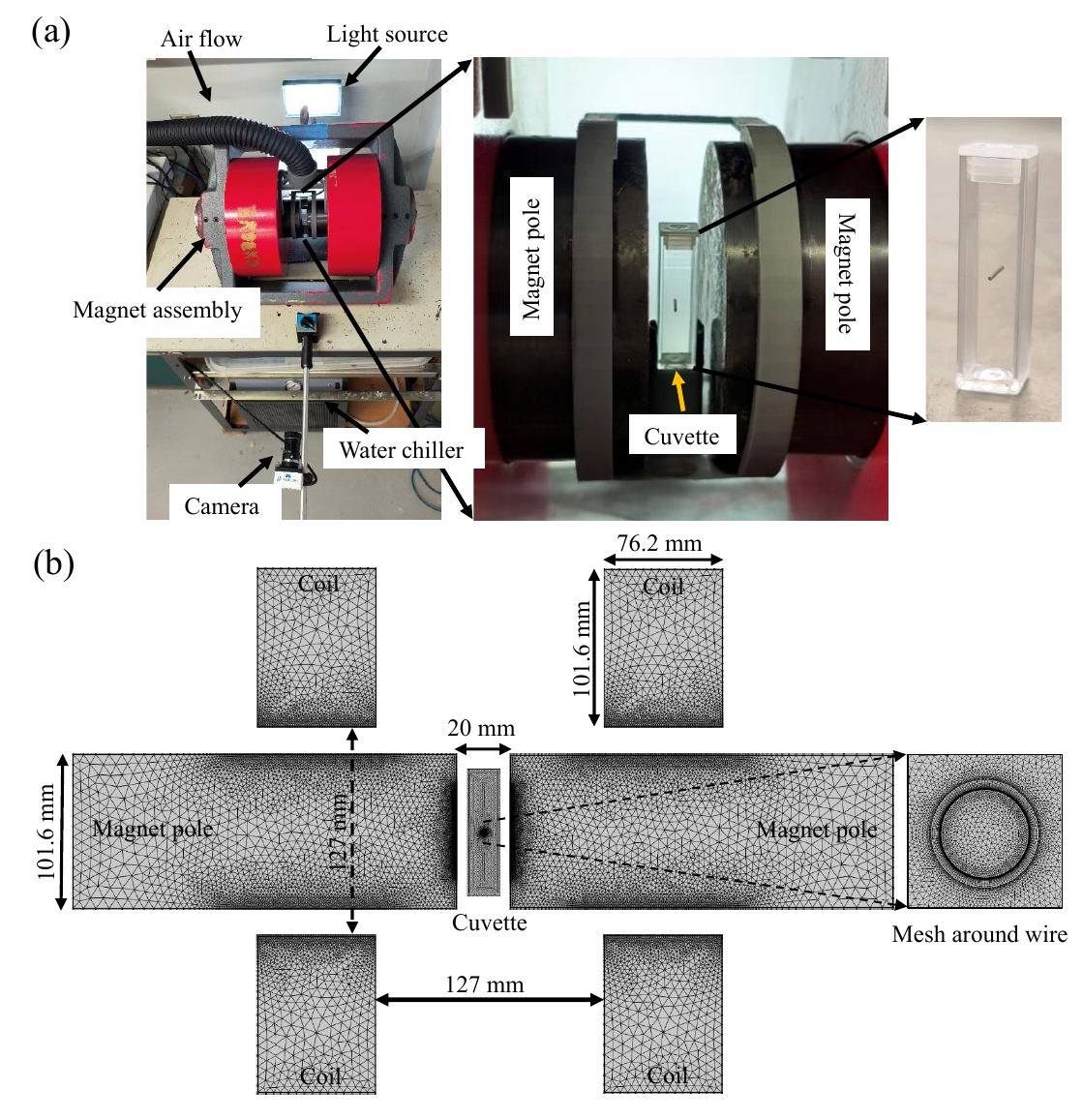}
 \caption{(a) The top view of the experimental setup consisting of the camera, light source, magnet and the temperature controller. Included in this part is also the closer view of the cuvette between the two flat poles. (b) A schematic depicting the two-dimensional representation of the electromagnet and the computational domain within the two poles with dimensions that are similar to the experimental setup.}
 \label{fig:ExpSetup}
\end{figure}  
\subsection{{\label{Method}Magnetophoresis setup}}
{The experimental setup for the magnetophoresis experiments comprises several key components. Figure~\ref{fig:ExpSetup}(a) shows the setup, where a 1T electromagnet, with poles approximately 10 cm in diameter, generates the magnetic field. The ferromagnetic wire is positioned centrally within the cuvette, perpendicular to the applied magnetic field. The wire is oriented with its long axis perpendicular to the applied magnetic field, ensuring that the field interacts symmetrically around its cylindrical surface, generating a radial field gradient. A camera is mounted to capture high-resolution images of the nano-particle dispersion around the wire and within the fluid domain. }\par 
\subsection{Concentration evaluation} To assess the spatio-temporal evolution of nano-particle concentration throughout the cuvette, we first measure the averaged light absorbance of the solution within the cuvette at various initial concentrations ranging from 10-100 mg/L. The light absorbance remains uniform during image acquisition. As shown in Fig.~S1 of the supplementary information, the normalized absorbance intensity within the cuvette increases linearly with initial concentration indicating that the Beer-Lambert law~\cite{bouguer1729essai,beer1852bestimmung} is valid (see the supplementary materials for further discussion on the calibration curves). This graph serves as a calibration curve that is used to evaluate the spatio-temporal evolution of particle concentration within the cuvette under the imposed external magnetic field. \HM{This calibration curve was obtained from solutions immediately after mixing. Consequently, potential changes in the calibration over the course of the experiment (typically two hours) were not accounted for. }
\section{\label{Methodology2}Multiphysics Simulations}
The multiphysics problem considered here requires the solution of three key dependent variables: the static magnetic field, momentum, and mass balance, which are discussed in details below. The multiphysics simulations are performed in two-dimension (2D), representing a cross-section of the cuvette and electromagnet because 3D simulation of the full apparatus are computationally cost-prohibitive. As shown in Figure~\ref{fig:ExpSetup}(b), the 2D simulation domain consists of a cuvette with dimensions similar to those used in experiments, and is placed between the magnet poles. To simulate the above system, a finite element-based numerical simulation technique was applied using COMSOL Multiphysics 6.1. The computational domain, which includes the surrounding environment with a radius of 2 m was discretized into $65548$ elements including $62508$ triangular meshes. Within this domain, the batch vessel was modeled by $11268$ elements with a $0.5$ mm maximum element size. The mesh component outside the batch vessel progressively grew in size with distance from the vessel center, with a maximum element growth rate $1.15$. \par

\subsection{{\label{Method}Static magnetic field}}
Before simulating the magnetophoresis of nano-particles, it is essential to simulate and validate the magnetic field distribution within the cuvette, especially around the stainless steel wire. To simulate the static magnetic field, we use Maxwell-Ampère's law:  
\begin{equation}
     \nabla \times \mathbf{H} = \mathbf{A};
     \label{Eq1}
\end{equation}
Where $\mathbf{H}$ and $\mathbf{A}$ are the strength of the magnetic field and the current density, respectively. Another important equation is the Gauss law and constitutive relation for the magnetization that allows us to determine the magnetic flux density within the domain: 
\begin{equation}
     \nabla \cdot \mathbf{B} = 0; \mathbf{B} = \mu_{0} \mu_{r} \mathbf{H}.
\end{equation}
Here $\mu_{0}$ is the permeability of free space, and $\mu_{r}$ is the relative permeability. 
\subsection{{\label{Method}Mass and momentum balance}}
{To evaluate the transport of nano-particles in the bulk fluid, we consider the force balance on the particle and fluid elements. 
The convective motion of the fluid element is governed by the continuity and Navier-Stokes equations. To account for the influence of a magnetic dipole, a magnetic force term is included, which affects both the particles and the surrounding fluid by transferring momentum as~\cite{huang1998thermoconvective}:
\begin{equation}
    \nabla \cdot  \mathbf{u_f} = 0.
\end{equation}
\vspace{-0.6cm}
\begin{equation}
    \rho \left[ \frac{\partial \mathbf{u_f}}{\partial t} + \mathbf{u_f}\cdot \nabla \mathbf{u_f}\right] =  \mu \nabla^2 \mathbf{u_f} + (\rho-\rho_l) \mathbf{g} + \frac{\chi_f}{\mu_{0}} (\mathbf{B} \cdot\nabla)\mathbf{B}.
    \label{NS}
\end{equation}
Here, $\mathbf{u_f}$, $\rho$, $\rho_l$, $\mathbf{g}$ and $\chi_f$ denote the fluid velocity, solution containing particle density, solvent density, gravitational acceleration and the volumetric magnetic susceptibility of the fluid element. In the case of a neutrally charged magnetic particle, its motion is governed by a force balance in the absence of inertia (Stokes' regime), which can be expressed as: \begin{equation}
 \mathbf{F}_{mp}+\mathbf{F}_{d} + \frac{4 \pi R^3_{p}}{3}(\rho-\rho_l) \mathbf{g}  = 0.
\label{ParticleForce}
\end{equation}
Here $\mathbf{F}_{mp}$ and $\mathbf{F}_d$ are the Kelvin and the drag force that particle experiences in the solution, which are expressed as: 
\begin{equation}
   \mathbf{F}_{mp} = \frac{4 \pi}{3} \frac{\Delta\chi R^3_{p} c}{\mu_{0}} (\mathbf{B} \cdot \nabla) \mathbf{B}; ~~~~~~~~~~~\mathbf{F}_{d} = -6 \pi \mu R_{p} \mathbf{u}_p.  
\end{equation}
In the above equations, $\Delta \chi$ is the molar magnetic susceptibility difference between particle and the surrounding fluid, $R_p$ is the particle radius, $c$ is the particle concentration in the fluid, and $\mathbf{u}_p$ is the magnetophoresis velocity of the particle relative to the fluid element.\par

The mass transport dynamics of the particles is given by a convective-diffusive equation expressed as:\begin{equation}
  \frac{\partial c}{\partial t} + \nabla \cdot \mathbf{N} = 0
    \label{MagCD}
\end{equation}
Here $\mathbf{N}$ is the total molar flux of particles presented as: $  \mathbf{N}  = -D \nabla c + c (\mathbf{u}_f+\mathbf{u}_p)$, with $D$ being the diffusion coefficient of the particles in the fluid. A no-slip boundary condition is set to all the walls of the cuvette and the wire surface. To account for the particle capture in a cuvette caused by magnetophoresis and sedimentation, the flux at the cuvette walls and wire surface, $\mathbf{N}_b$, was defined as follows:
\begin{equation}
    \mathbf{N}_b = \left [\frac{2 R^2_{p} \Delta \chi}{9 \mu_{0} \eta } c (\mathbf{B}. \nabla) \mathbf{B} + \frac{2 R^2_{p} \nabla \rho}{9 \eta} c \mathbf{g} \right].\mathbf{n} 
\end{equation}
Where $\mathbf{n}$ is the normal outward vector at the boundary. The flow in the domain assumed to be inertialess because the particle Reynolds number is calculated to be $Re << 1$. By solving Eq.~(\ref{Eq1}-\ref{MagCD}), in the computational domain, we can determine the magnetophoretic behavior of the particles around the wire and results will be compared with the experimental observations.
}\par
\begin{figure}[hthp]
 \centering
 \includegraphics[trim=0cm 12.5cm 0.5cm 0cm,clip,width=13.5cm]{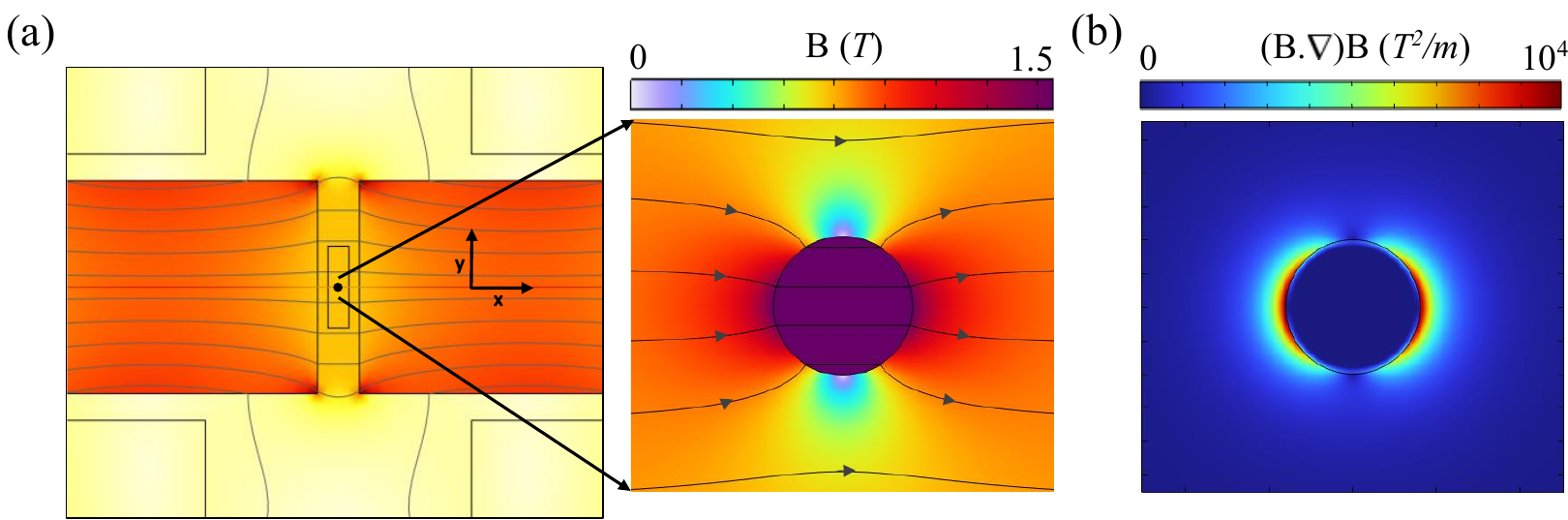}
 \includegraphics[trim=0.35cm 7cm 0cm 0cm,clip,width=14cm]{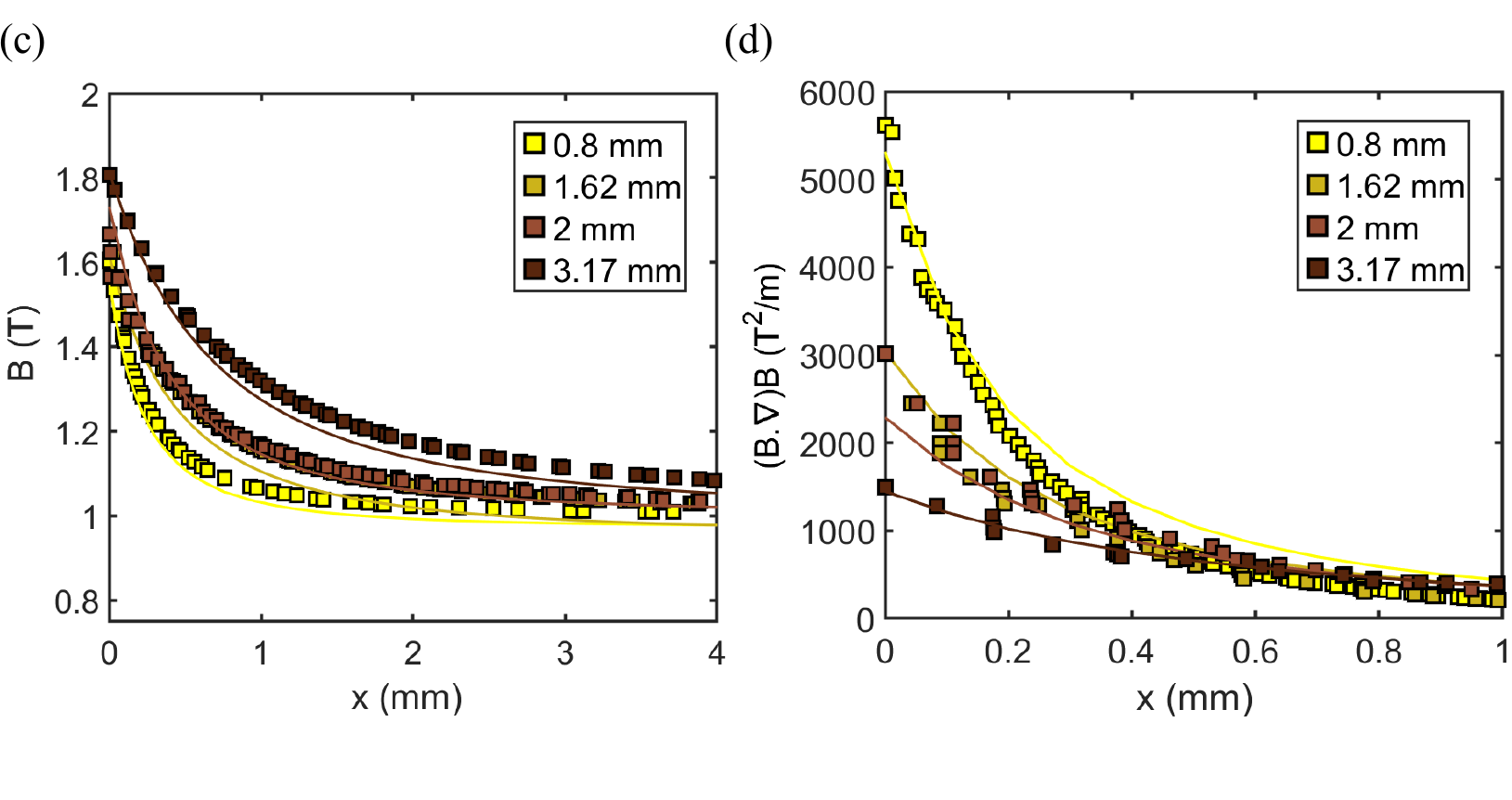}
 \caption{(a) The 2D simulated magnetic flux density around the wire with a diameter of 0.8 $mm$ under influence of a $1$ $T$ magnetic field strength. The arrow represents the direction of the magnetic field. (b) The 2D magnetic field gradients profile around the wire under the same conditions as in part (a). (c) Averaged magnetic flux density and magnetic flux density gradients (d) around wires of different diameters at an applied magnetic field of $1$ T. In (c,d) symbols represent the results of simulations and the curves represent the theoretical predictions. Here x = 0 represents the surface of the wire. } 
 \label{fig:MagField}
\end{figure}

\vspace{-0.5cm}
{\section{{\label{Results}Results and Discussions}}

\subsection{Static Magnetic Field}

As a first step, we computed the magnetic flux density around the wire. Fig.~(\ref{fig:MagField}) illustrates the spatial distribution of the simulated magnetic flux density ($\mathbf{B}$) around the wire. In addition,  Fig.~\ref{fig:MagField}(b) shows the magnetic field gradients around the wire. To ensure the validity of the simulated magnetic field around the wire, one should compare the calculations to experimentally measured data. However, because of the highly localized nature of the magnetic flux density near the wire surface, experimental measurements of the magnetic flux density in this region are challenging. Therefore, to validate the simulation results, we compare them to the closed-form analytical solution around a single wire suggested in prior published literature~\cite{svoboda2004magnetic} (The equations used to analytically solve for the magnetic field and its gradient around the wire are provided in the supplementary materials.). Fig.~\ref{fig:MagField} (c,d) shows the calculated magnetic flux density and gradients as a function of distance from the wire surface (symbols) along with the analytical solutions (continuous curves). The strong agreement between the simulated and analytical results validates the accuracy of our magnetic field calculations, confirming their reliability for use in subsequent magnetophoresis simulations.} 
\subsection{Particle sedimentation}

We first examine the simplified case of particle transport under gravity in the absence of an external magnetic field. {Figure~\ref{fig:B0}(a,c) shows the spatio-temporal evolution of both paramagnetic and diamagnetic particles in the absence of an external magnetic field measured in experiments (top row) and calculated from simulations (bottom row). Note that, as a result of particle sedimentation, the particle concentration along the longest axis of the cuvette varies, with fewer particles at the top and more at the bottom. To evaluate the overall particle depletion within the domain, we define an averaged normalized concentration as ${\langle C \rangle}/{C_0}$, where $\langle C \rangle$ represents the surface-averaged particle concentration at any given time, and $C_0$ is the initial particle concentration. Figure~\ref{fig:B0}(b,d) show the temporal evolution of the averaged normalized concentration for various initial concentrations. Over time, particle depletion within the cuvette leads to a reduction in particle concentration. Additionally, higher initial particle concentrations correspond to an increased rate of depletion. However, the sedimentation rate remains low, with a maximum of 2\% of paramagnetic particles of MnO$_2$ and 4\% of diamagnetic particles of Bi$_2$O$_3$ being removed over a 2-hour period. The particle supplier specified radii of 50 nm and 40 nm for MnO$_2$ and Bi$_2$O$_3$ particles, respectively. To assess the relative importance of the gravity to thermal diffusion, one can construct a gravitational Peclet number as: 
\begin{figure}[hthp]
 \centering
 \includegraphics[trim=0cm 9.5cm 0cm 0cm,clip,width=16cm]{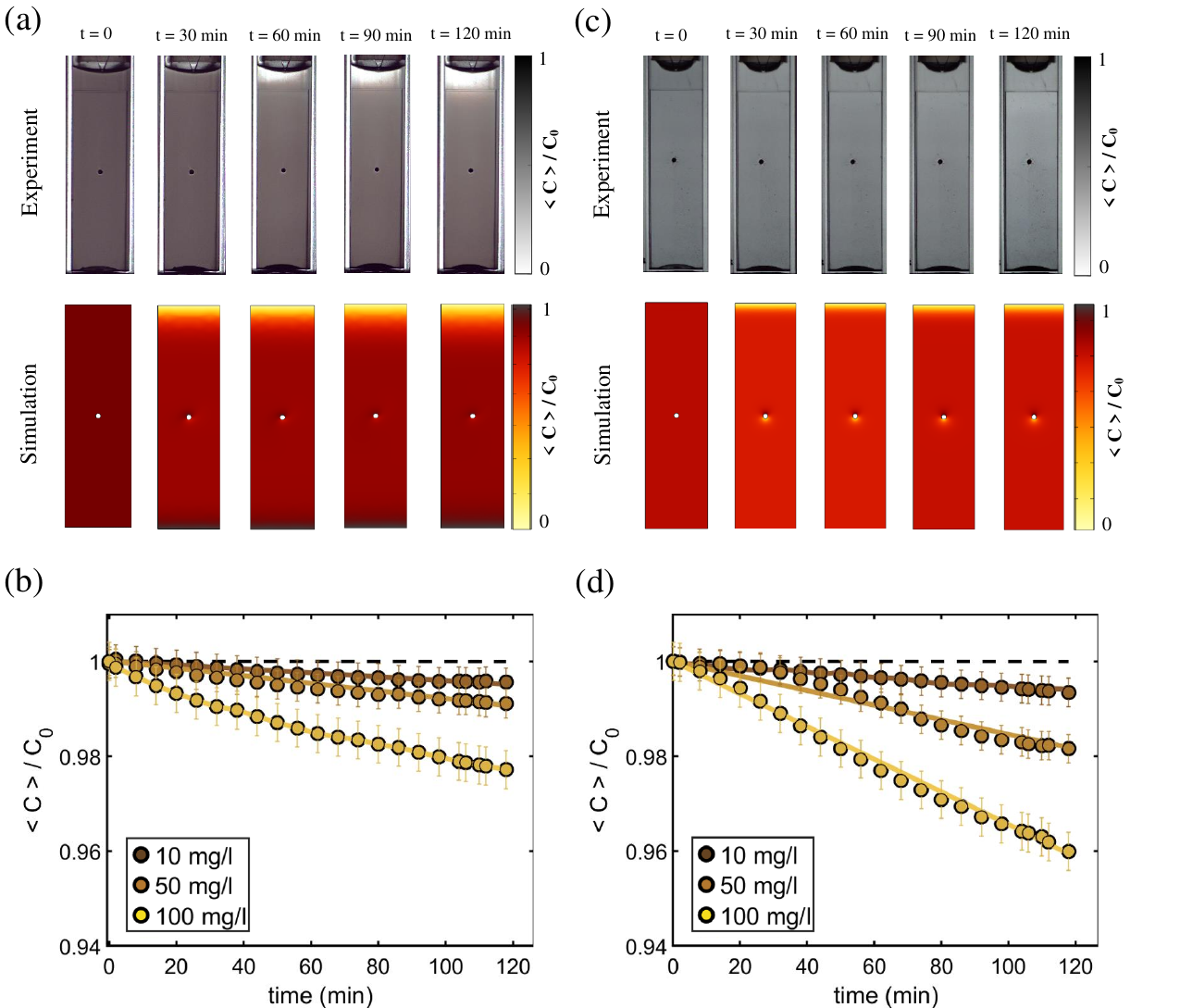}
 \caption{{Control experiments to assess particle sedimentation rate. The spatio-temporal evolution of (a) MnO$_2$, and (b) Bi$_2$O$_3$ nano-particle concentrations as measured in experiments at an initial concentration of 100 [mg/L], magnetic field $\mathbf{B}$ = 0 [T] and wire diameter $d$ = 0.8 [mm] (top row) and calculated through numerical simulations (bottom row). (b,d) The averaged normalized concentration corresponding to part (a,c) for experimental data (symbols) and numerical simulation results (lines)  for different initial concentrations. Note that the dashed line denotes the simulation results for vendor provided nano-particle size. }}
 \label{fig:B0}
\end{figure}
\begin{equation}
    \mathrm{Pe_g} = \frac{4 \pi R^4_{p}(\rho-\rho_l) \mathbf{g} }{3k_B T}.
    \label{eqn:Peclet-definition}
\end{equation}
For those particle sizes provided by the vendor, $\mathrm{Pe_g} \approx \mathcal{O}({10^{-5}})$, which suggests that sedimentation should be minimal. However, initial simulations using these sizes significantly underpredicted the spatio-temporal evolution of particle concentrations (shown as dashed lines in Fig.~\ref{fig:B0}(b,d)). To address this discrepancy, we used a particle size distribution with three species in our simulations as $\sum_{i=1}^{3} \phi_i R_{pi}$, where $\phi_i$ and $R_{pi}$ are the particle mass fraction and its radius. For this purpose, we considered up to three particle size classes, as including more classes would result in extremely high computational costs. The particle size and their mass fraction were varied to produce the best match with experimental data as shown as continuous curves in Fig.~\ref{fig:B0}(b,d)). Table~(\ref{particlesizeB0}) shows the particle size distribution that generated the best match with experiments. With increasing initial particle concentration, the particle size distribution shifts toward larger particles. This discrepancy between the calculated particle sizes and those provided by the vendor is presumably due to the formation of particle aggregates. Nevertheless, accurate calculations of particle size and size distribution are essential for reliable subsequent magnetophoresis simulations.} \HM{To independently assess the particle size distribution under no magnetic field, we conducted dynamic light scattering (DLS) measurements. While DLS is inherently more sensitive to larger particles due to its intensity-weighted nature, the measured hydrodynamic radius was $R_p \approx 320 \pm 20 \mathrm{nm}$ for MnO$_2$ and $R_p \approx 330 \pm 30 \mathrm{nm}$ for Bi$_2$O$_3$ with a polydispersity index of approximately 0.3. The particle size distribution inferred from simulations are consistent with the DLS results in indicating the presence of larger particles.} 
\begin{table}[h!]

\centering
\caption{\centering Particle size distribution for simulations performed for $\textbf{B} = 0$ T condition.}
\begin{tabular}{|c|c c c| c c c| c c c| c c c|}
\hline
{C$_0$} [mg/l] & \multicolumn{6}{c|}{\textbf{MnO$_2$}} & \multicolumn{6}{c|}{\textbf{Bi$_2$O$_3$}} \\ \cline{2-13} 
 & \multicolumn{3}{c|}{$\phi_i$} & \multicolumn{3}{c|}{R$_{pi}$ [nm]} & \multicolumn{3}{c|}{$\phi_i$} & \multicolumn{3}{c|}{R$_{pi}$ [nm]} \\ \hline
10   & 0.82 & 0.1 & 0.08 & 50 & 100 & 150 & 0.84 & 0.1 & 0.06 & 40 & 80 & 120\\ 
50  & 0.75 & 0.15 & 0.10 & 50 & 150 & 200 & 0.80 & 0.12 & 0.08 & 40 & 80 & 120 \\ 
100 & 0.75 & 0.15 & 0.10 & 100 & 200 & 250 & 0.75 & 0.15 & 0.10 & 80 & 120 & 160 \\ \hline
\end{tabular}
\label{particlesizeB0}
\end{table}
{\subsection{Magnetophoresis}
\subsubsection{Initial particle concentration}Following experiments conducted without a magnetic field, a 1 T external magnetic field was applied to the solution of manganese oxide nanoparticles at various initial concentrations. Figure~\ref{fig:Fig2}(a) illustrates the spatiotemporal evolution of particle concentration in the cuvette, with the top row representing experimental results. As shown in Fig.~\ref{fig:Fig2}(a), and shortly after the magnetic field is applied, particles begin to move rapidly from the top and bottom of the wire to the sides (left and right), where the magnetic field gradient is strongest (see also Movie 1 in the supplementary materials). The motion of particles transfers the momentum to the fluids and creates two symmetrical secondary vortices near the wire, causing the fluid to rise in the wake of the wire. As particles move toward the wire, the surrounding liquid becomes relatively less dense, creating a buoyant force that drives the lighter, depleted liquid upward. Over time, this secondary convective flow significantly enhances particle transport across the domain, concentrating particles along the flanks of the wires where the magnetic field gradient is highest. Consequently, the overall particle concentration within the cuvette decreases, as most particles are captured by the wire. } 
\begin{figure}[hthp]
 \centering
 \includegraphics[trim=0cm 0cm 4cm 0cm,clip,width=10cm]{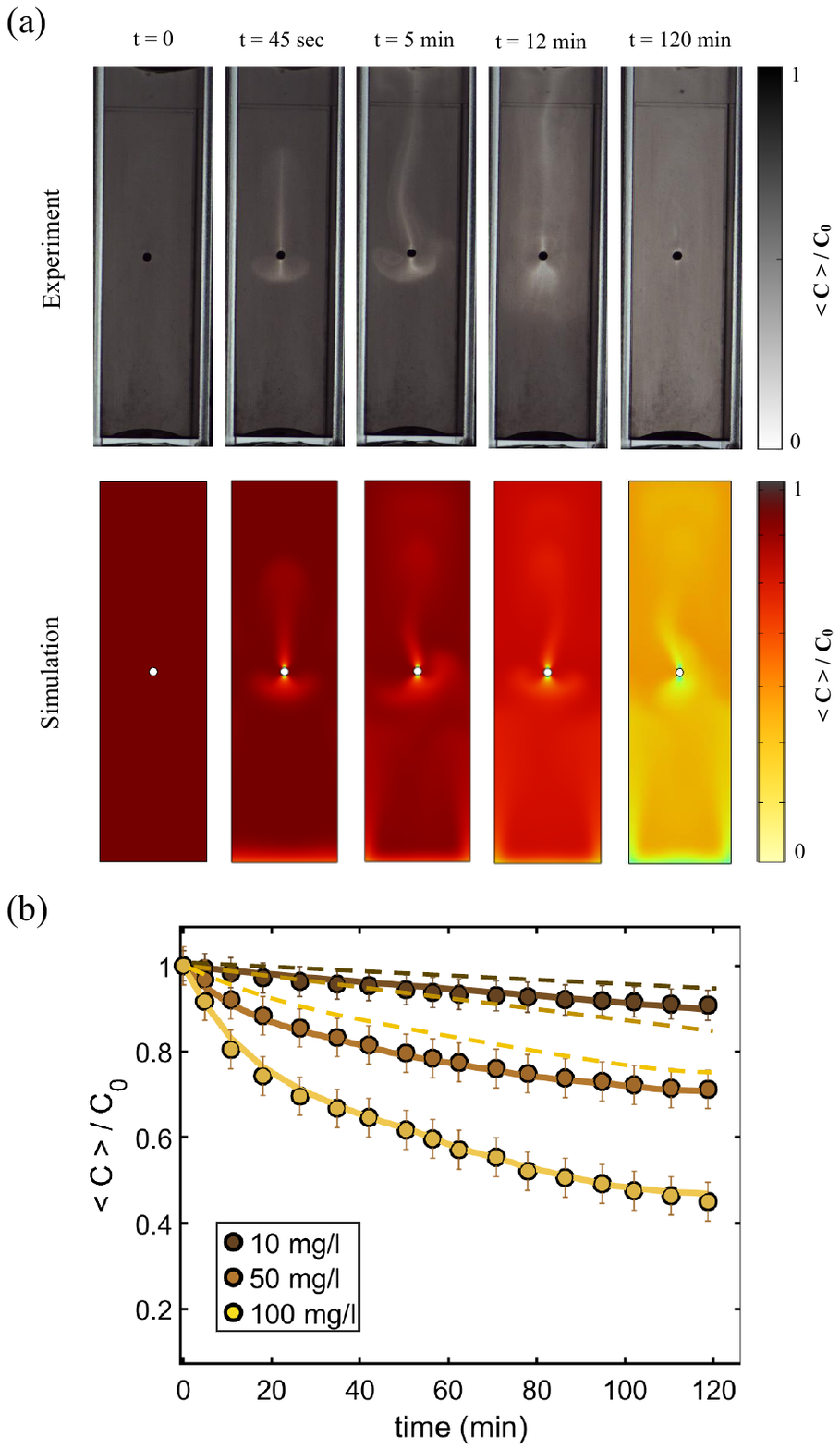}
\caption{(a) Spatio-temporal evolution of particle concentrations measured in experiments (top row), and calculated (bottom row) for manganese oxide particle at a concentration = 100 mg/L, and a magnetic field $\mathbf{B}$ = 1 T and a wire diameter $d$ = 0.8 mm. (b) Volume-averaged normalized concentration as a function of time for experiments (symbols), and numerical simulations. The dashed lines are simulations with particle size distribution from Table~(\ref{particlesizeB0}) and the continuous curves are simulations that best match the experiments with particle size distribution shown in Table~(\ref{particlesizeB1}).}
 \label{fig:Fig2}
\end{figure}
These experiments suggest that magnetophoresis of this weakly magnetic nano-particle is much stronger than the sedimentation dynamics. The latter result is perhaps not surprising, since the particles experience an additional force under the influence of a magnetic field. To assess the relative significance of this force, a magnetic Peclet number is defined. The magnetic Peclet number, which represents the ratio of magnetic force to diffusion is defined as:
\begin{equation}
    \mathrm{Pe_m} = \frac{\mathbf{F}_{mp}}{\mathbf{F}_{T}}= \frac{4 \pi}{3} \frac{\Delta\chi R^4_{p} c}{\mu_{0}k_B T} (\mathbf{B} \cdot \nabla) \mathbf{B}.
    \label{eqn:Peclet-definition}
\end{equation}
The magnetic Peclet number varies across the domain due to the significant variation in the magnetic field gradient. Thus, we define a maximum magnetic Peclet number within the domain as:
\begin{equation}
    \mathrm{Pe_m}|_{max} = \frac{\mathbf{F}_{mp}}{\mathbf{F}_{T}}= \frac{4 \pi}{3} \frac{\Delta\chi R^4_{p} c}{\mu_{0}k_B T} (\mathbf{B} \cdot \nabla) \mathbf{B}|_{max}.
    \label{eqn:Peclet-definition}
\end{equation}
Our calculations show that the ratio of the magnetic Peclet number to the gravitational Peclet number is significantly large, $\mathrm{Pe_m}|_{max} / \mathrm{Pe_g} \gg 1$. This confirms that the magnetic force around the wire overwhelmingly dominates the gravitational effects. As a result, nano-particle transport is primarily governed by magnetophoresis, leading to rapid particle capture (or depletion) within the domain under the applied magnetic field. \par

Figure~\ref{fig:Fig2}(b) shows the temporal evolution of the average normalized manganese oxide nano-particle concentration for various initial concentrations. As the initial particle concentration increases, the rate of particle depletion measured within the cuvette increases. This result is different from the results of Leong et al.\cite{leong2015magnetophoresis} and Rassolov et al~\cite{rassolov2025} who showed that under a low magnetic field gradient, the temporal evolution of normalized concentration does not depend on the initial concentration of nano-particles. To gain a more in-depth understanding of these experimental results, we conducted multiphysics numerical simulations of particle transport under an applied magnetic field of 1T. Included in Figure~\ref{fig:Fig2}(b) are also the multiphysics simulation results. The simulations were first attempted with the particle size distribution noted in Table~(\ref{particlesizeB0}), and are shown as dashed curves in Fig.~\ref{fig:Fig2}(b) at various initial concentrations. Although these simulations predict faster particle depletion at higher concentrations compared to no magnetic field experiments, they significantly underestimate those concentration variation levels observed in experiments. To better align simulations with experimental observations, we optimized the particle size distribution to achieve the best match with the experimental results. The sample snapshots displaying the simulation results that best match the experiments are shown in the bottom row of  Figure~\ref{fig:Fig2}(a), and as continuous curves in Fig.~\ref{fig:Fig2}(b). The resulting particle size distribution is also presented in Table~(\ref{particlesizeB1}). First, as shown in the bottom row of Figure~\ref{fig:Fig2}(a), the simulations capture the overall experimental observations on the formation of secondary vortices near the wire and the motion of particles upward above the wire (see the velocity vectors in the Fig.~S2(a) of the supplementary materials). Secondly, in order to match the results of simulations with experiments, the particle size or the mass fraction of larger particles within the distribution needs to be increased (c.f., particle size distribution of manganese oxide of Table~\ref{particlesizeB1} and Table~\ref{particlesizeB0}). The latter result suggests that the paramagnetic nanoparticles may have undergone magnetic field-induced cluster formation in the experiments. We will return to this point towards the end of the manuscript.\par
\begin{figure}[hthp]
 \centering
 \includegraphics[trim=0cm 0cm 4cm 0cm,clip,width=11cm]{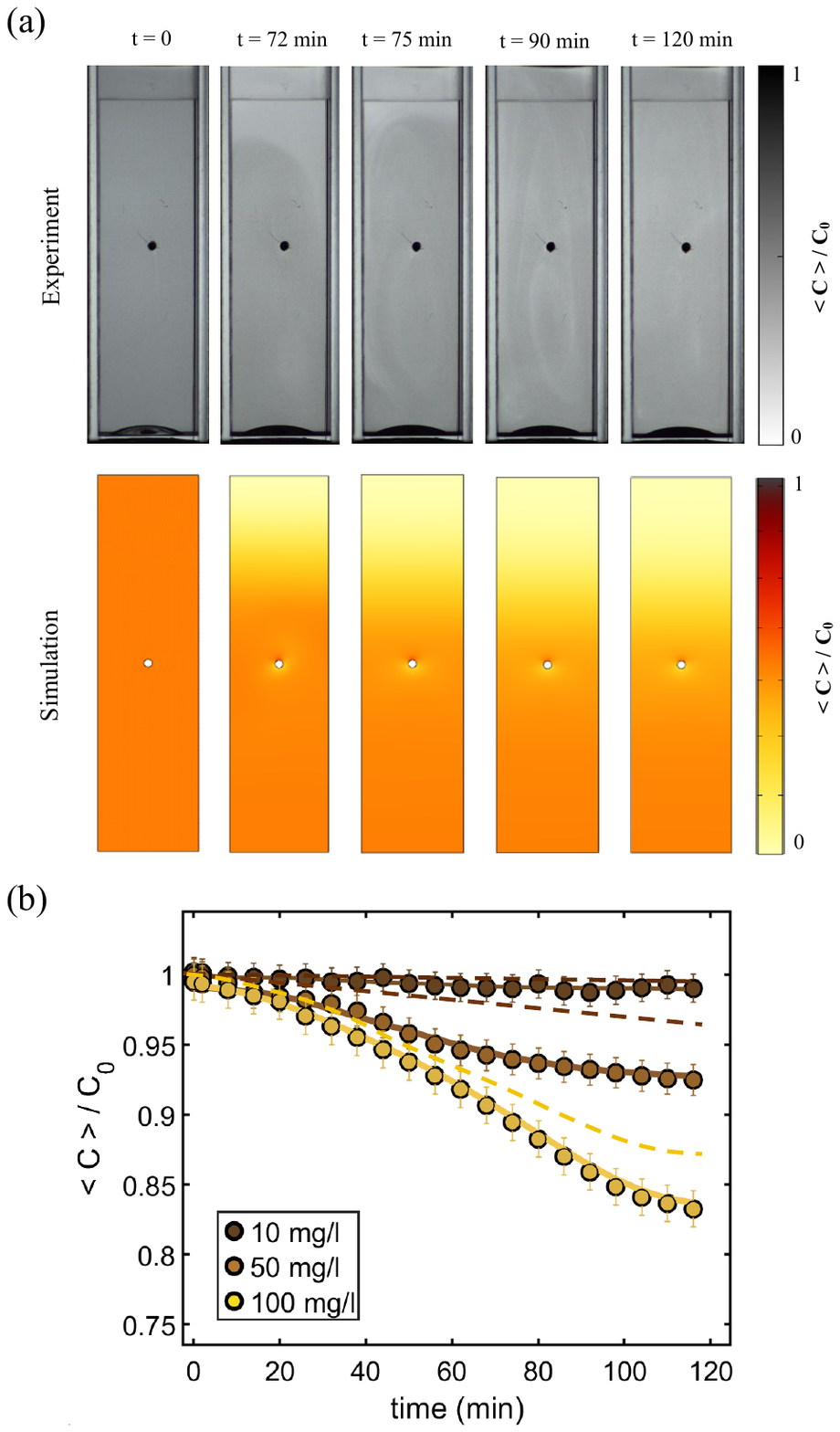}
   \caption{(a) The spatio-temporal evolution of bismuth oxide concentration measured in experiments (top row) and calculated in simulations (bottom row) at a concentration of 100 mg/L, a magnetic field of $\mathbf{B}$ =1 T, and a wire diameter of $d$ = 0.8 mm. (b) Averaged normalized concentration of diamagnetic particles as a function of time showing experimental data (symbols) with simulation results (curves). The dashed lines are simulations with particle size distributions from Table~(\ref{particlesizeB0}) and the continuous curves are simulations that best match the experiments, and the corresponding particle sizes shown in Table~(\ref{particlesizeB1}). }   
 \label{fig:Fig3}
\end{figure}

 
\begin{table}
\centering 
\caption{\centering Particle size distribution for simulations performed under external magnetic field of $\mathbf{B}$ = 1 T.}
\begin{tabular}{|c|c c c| c c c| c c c| c c c|}
\hline
\textbf{C$_0$} [mg/l] & \multicolumn{6}{c|}{\textbf{MnO$_2$}} & \multicolumn{6}{c|}{\textbf{Bi$_2$O$_3$}} \\ \cline{2-13} 
 & \multicolumn{3}{c|}{$\phi_i$} & \multicolumn{3}{c|}{R$_{pi}$ [nm]} & \multicolumn{3}{c|}{$\phi_i$} & \multicolumn{3}{c|}{R$_{pi}$ [nm]} \\ \hline
10 & 0.75 & 0.15 & 0.10 & 50 & 150 & 200 & 0.82 & 0.10 & 0.08 & 40 & 80 & 120 \\ 
50  & 0.75 & 0.15 & 0.10 & 100 & 300 & 500 & 0.75 & 0.15 & 0.10 & 40 & 80 & 120 \\ 
100 & 0.70 & 0.20 & 0.10 & 150 & 500 & 800 & 0.70 & 0.17 & 0.13 & 80 & 120 & 160 \\ \hline
\end{tabular}
\label{particlesizeB1}
\end{table}
In addition to paramagnetic nanoparticles, we conducted magnetophoresis experiments using diamagnetic bismuth oxide nanoparticles, as shown in Fig.~(\ref{fig:Fig3}). Unlike the manganese oxide nanoparticles, secondary vortices were not readily observed. For bismuth oxide, the magnetic force induced a subtle secondary flow, which became discernible approximately one hour after the application of the external magnetic field (see Movie 2 in the supplementary materials). 
The results for bismuth oxide nanoparticles reveal several notable differences between experiments conducted with and without the application of a magnetic field. Under the influence of the magnetic field, bismuth oxide particles deplete from the cuvette more rapidly compared to when no magnetic field is present. Furthermore, as the initial concentration of bismuth oxide nanoparticles increases, the rate of particle depletion also increases. Observations indicate that, under the magnetic field, particles are deposited both at the bottom of the cuvette and at the top of the wire, where the magnetic field gradient weakens. In contrast, in the absence of a magnetic field, the particles settle predominantly at the bottom of the cuvette. A second key difference lies in the temporal evolution of the average particle concentration. At higher concentrations, experiments with the magnetic field show an initially slower rate of particle depletion, which then accelerates over time before leveling off at longer times. This faster depletion under the magnetic field is likely due to magnetic repulsion, which drives the diamagnetic particles away from the flanks of the wire and mixes them more effectively within the domain. Additionally, particles are attracted toward the top of the wire where the magnetic field gradient diminishes, as illustrated in Fig.~\ref{fig:MagField} and further supported by the velocity profiles shown in Fig.~S2(b) of the supplementary materials.\par 

Numerical simulations using the particle size distribution of bismuth oxide (Table~\ref{particlesizeB0}) are presented in Fig.~\ref{fig:Fig3} as dashed curves. The simulations that best match the experimental results are shown in the bottom row of Fig.~\ref{fig:Fig3}(a), representing the spatio-temporal evolution of particle concentration, and as continuous curves in Fig.~\ref{fig:Fig3}(b). Interestingly, these simulations suggest that achieving the best agreement with the experimental data requires an increase in the volume fraction of larger bismuth oxide nanoparticles (c.f., bismuth oxide particle size distribution of Table~\ref{particlesizeB0} and Table~\ref{particlesizeB1}). This finding implies that diamagnetic nanoparticles may potentially form larger aggregates under the influence of the magnetic field, though to a significantly smaller extent than paramagnetic particles. To the best of our knowledge, field-induced clustering of diamagnetic nanoparticles has not been reported previously. We will revisit this observation in greater detail later in the manuscript.



{\subsubsection{Magnetic field strength} To further confirm the effects of magnetic field on the magnetophoresis of these nanoparticles, we performed experiments at a fixed initial concentration and different magnetic field strengths. 
Fig.~\ref{fig:Fig8}(a-c) shows the effect of the external magnetic field strength on the magnetophoretic dynamics of both paramagnetic and diamagnetic nanoparticle suspensions. For both paramagnetic and diamagnetic particles, increasing the magnetic field strength, increases the rate of particle depletion from the cuvette. As the magnetic field is strengthened, the magnetic field gradient around the wire is expected to increase. As a result, both particles experience larger magnetic forces and enhanced magnetophoretic effects. It is worth noting that the variation in concentration at a given time is a strong function of the applied magnetic field. As shown in Fig.~\ref{fig:Fig8}(c), at weak magnetic fields (e.g., 0.25 T), the paramagnetic and diamagnetic particles exhibit minimal magnetophoresis. However, at 0.5 T, there is a marked increase in the rate of paramagnetic nanoparticle magnetophoresis. For diamagnetic particles, magnetophoresis remains a weak function of the magnetic field up to 0.75 T, after which, at 1 T, the rate of magnetophoresis increases dramatically. 
\begin{figure} [hthp]
 \centering
 \includegraphics[trim=0cm 14cm 0cm 0cm,clip,width=17cm]{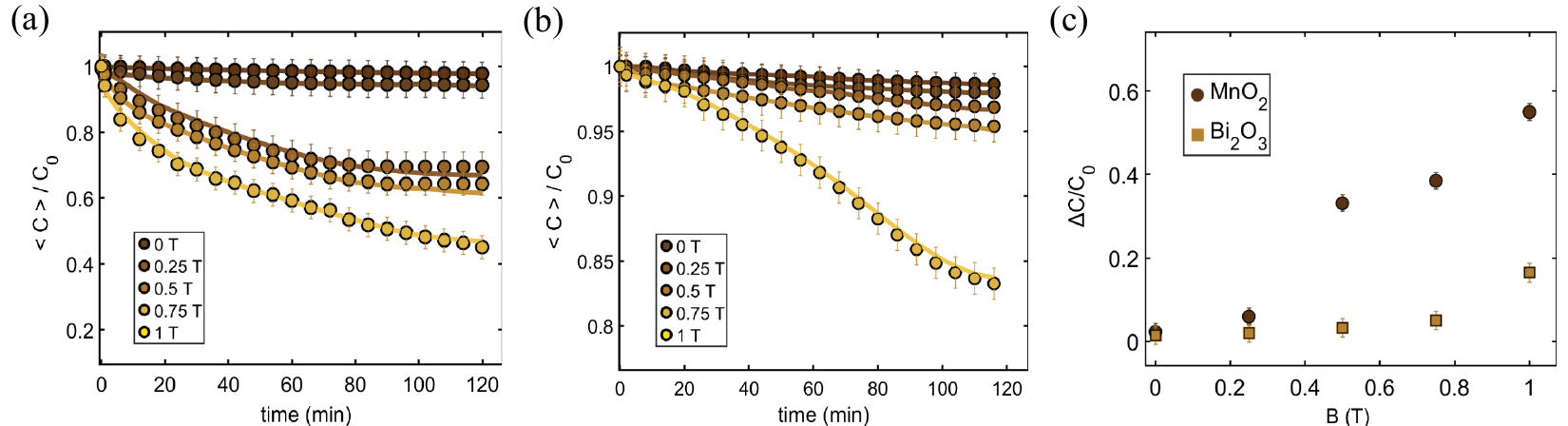}
 \caption{Temporal evolution of normalized particle concentrations for paramagnetic (a), and diamagnetic (b) particles as a function of magnetic field strength with an initial concentration of 100 mg/mL and a wire diameter $d$ = 0.8 mm. In (a,b) symbols represent the experimental data, and the curves represent the simulation results that best match the experiments. (c) The normalized concentration change as a function of imposed magnetic field after a 2-hour period both for paramagnetic (circles) and diamagnetic particles (squares). }
 \label{fig:Fig8}
\end{figure}
Table~(\ref{particlesize_B}) presents the particle size distributions of the simulations that are best aligned with the experimental data (also shown as continuous lines in Fig.~\ref{fig:Fig8}(a,b)). Notably, as the external magnetic field strength increases, the size distribution of paramagnetic MnO$_2$ shifts toward larger particles with a higher mass fraction of larger particles. This trend suggests that MnO$_2$ particles undergo field-induced clustering. In contrast, the diamagnetic Bi$_2$O$_3$ particles exhibit only a slight shift in their size distribution across the tested magnetic field range, indicating a weaker or negligible clustering effect.

\begin{table}[h!]
\centering
\caption{\centering Particle size distribution for different magnetic fields $B$ obtained from simulations performed at an initial concentration of 100 mg/L and a wire diameter $d$ = 0.8 mm.}

\begin{tabular}{|c|c c c| c c c| c c c| c c c|}
\hline
\textbf{Magnetic field} [T] & \multicolumn{6}{c|}{\textbf{MnO$_2$}} & \multicolumn{6}{c|}{\textbf{Bi$_2$O$_3$}} \\ \cline{2-13} 
 & \multicolumn{3}{c|}{$\phi_i$} & \multicolumn{3}{c|}{R$_{pi}$ [nm]} & \multicolumn{3}{c|}{$\phi_i$} & \multicolumn{3}{c|}{R$_{pi}$ [nm]} \\ \hline
0 & 0.75 & 0.15 & 0.10 & 100 & 200 & 250 & 0.75 & 0.15 & 0.10 & 80 & 120 & 160 \\ 
0.25  & 0.74 & 0.16 & 0.10 & 100 & 250 & 300 & 0.74 & 0.16 & 0.10 & 80 & 120 & 160 \\ 
0.5 & 0.72 & 0.17 & 0.11 & 100 & 350 & 500 & 0.73 & 0.16 & 0.11 & 80 & 120 & 160 \\ 
0.75  & 0.72 & 0.17 & 0.11 & 150 & 400 & 600 & 0.72 & 0.17 & 0.11 & 80 & 120 & 160 \\ 
1 & 0.70 & 0.20 & 0.10 & 150 & 500 & 800 & 0.70 & 0.17 & 0.13 & 80 & 120 & 160 \\ \hline
\end{tabular}
\label{particlesize_B}
\end{table}
\subsubsection{Wire diameter}
The effect of the diameter of the wire on the magnetophoresis of the particles is presented in Fig.~\ref{fig:Fig7}(a,b) in terms of the temporal evolution of normalized concentration. For paramagnetic particles, increasing the wire diameter enhances magnetophoresis dynamics significantly, leading to a greater particle depletion from the cuvette for larger wires (see Fig.~\ref{fig:Fig7}(a), and Fig.~S3 for the corresponding spatio-temporal evolution of particle concentration, and Movie 3 in the supplementary materials). As the wire diameter increases, two key factors influence magnetophoretic capture. First, as shown in Fig.~\ref{fig:MagField}, the maximum magnetic field gradient near the wire surface is inversely proportional to the wire diameter, meaning that larger wires generate weaker local magnetic field gradients. Secondly, a larger wire diameter provides a greater overall surface area for magnetic particle capture, potentially enhancing collection efficiency. These competing effects determine the net capture rate, and our experimental results suggest that the increase in surface area dominates, leading to enhanced particle capture with larger wires. To further investigate the influence of the wire diameter on magnetophoresis, we compared our experimental data with numerical simulations in Fig.~\ref{fig:Fig7}(a,b). Notably, the simulations reveal that as the wire diameter increases, the particle size distribution shifts toward a higher proportion of larger particles and fewer smaller ones (see Table~(\ref{particlesize_d})). If the same particle size distribution obtained for a 0.8 mm wire were applied to larger wires, the simulations would underpredict the observed concentration depletion. This discrepancy suggests that that increased surface area of larger wires may facilitate enhanced cluster formation due to an expanded magnetically induced capture zone, further contributing to the observed increase in particle retention.\par 

\begin{figure}
 \centering
 \includegraphics[trim=0cm 19cm 2.5cm 0cm,clip,width=16cm]{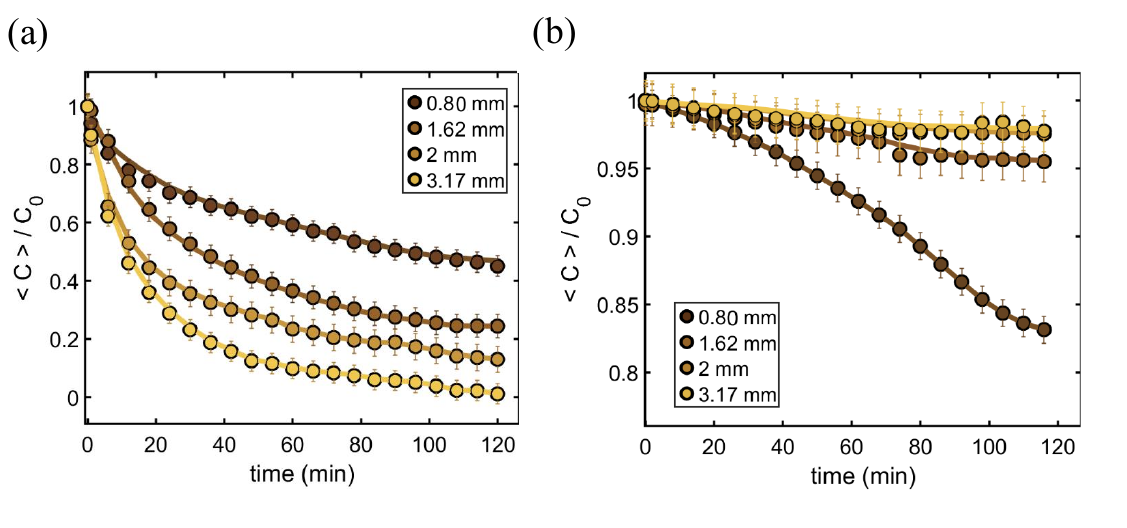}
 \caption{The temporal evolution of normalized averaged concentration in the cuvette for paramagnetic particles (a) and diamagnetic particles (b) for different wire diameters, and at an initial concentration of 100 mg/L under a magnetic field of $\mathbf{B}=$ 1T. The markers represent the experimental data, and the curves represent the simulation results that best match experiments. }
 \label{fig:Fig7}
\end{figure}
Figure~\ref{fig:Fig7}(b) presents the temporal evolution of the average concentration for diamagnetic particles, revealing a striking contrast to the behavior observed for paramagnetic manganese oxide particles (see also Fig.~S4 in the supplementary materials). As the wire diameter increases, the rate of concentration depletion within the domain slows significantly. Notably, for larger wire diameters, the depletion rate becomes even lower than in the absence of a magnetic field. This counterintuitive trend arises from the inherently repulsive nature of the magnetic force along the wire's flanks acting on diamagnetic particles. For smaller wires, the stronger localized magnetic field gradients generate a greater repulsive force from the sides of the wire while simultaneously exerting an attractive force toward the top. When combined with gravitational settling, this results in a larger net force that enhances particle sedimentation and accumulation along the top side of the wire within the cuvette. In contrast, as the wire diameter increases, the magnetic field gradient weakens, thereby reducing both the repulsive and attractive forces, ultimately diminishing the overall particle transport effect. Our numerical simulations further support this finding, showing that the particle size distribution of bismuth oxide remains largely unchanged for larger wires compared to control measurements performed without a magnetic field. \par 

\begin{table}[h!]
\centering
\caption{\centering Particle size distribution for different wire diameters obtained from simulations performed under an external magnetic field of $\textbf{B}$ = 1 T at an initial concentration of 100 mg/L.}
\begin{tabular}{|c|c c c| c c c| c c c| c c c|}
\hline
\textbf{Wire diameter} [mm] & \multicolumn{6}{c|}{\textbf{MnO$_2$}} & \multicolumn{6}{c|}{\textbf{Bi$_2$O$_3$}} \\ \cline{2-13} 
 & \multicolumn{3}{c|}{$\phi_i$} & \multicolumn{3}{c|}{R$_{pi}$ [nm]} & \multicolumn{3}{c|}{$\phi_i$} & \multicolumn{3}{c|}{R$_{pi}$ [nm]} \\ \hline
1.62 & 0.70 & 0.20 & 0.10 & 150 & 550 & 800 & 0.75 & 0.15 & 0.10 & 80 & 120 & 160 \\ 
2.0  & 0.70 & 0.20 & 0.10 & 200 & 600 & 800 & 0.78 & 0.12 & 0.10 & 80 & 120 & 160 \\ 
3.17 & 0.65 & 0.20 & 0.15 & 200 & 600 & 800 & 0.79 & 0.12 & 0.09 & 80 & 120 & 160 \\ \hline
\end{tabular}
\label{particlesize_d}
\end{table}

{\subsubsection{Multiple wires} 
Since magnetophoresis is driven by the magnetic field gradient around the magnetized wire, it is reasonable to expect that increasing the number of wires would enhance the magnetophoretic effect on these nanoparticles. In this part, we investigate how the number of wires influences the magnetophoresis of these particles. To this end, we selected the largest wire size and varied the number of wires to two and three. Fig.~(\ref{fig:Fig9}) shows the spatiotemporal evolution of particle concentration for two wire configurations, and for the three wire configuration is presented in Fig.~S5 in the supplementary materials. When a magnetic field is applied, the magnetophoresis of particles starts in a manner similar to that observed before for a single wire configuration. However, the presence of an additional wire in a cuvette causes a significant change in the fluid flow pattern around the wires. While secondary flows are observed at the beginning near each wire, soon they interact with each other and this interaction suppresses the structure of each vortex. The mixing of these vortices led to a slight acceleration of paramagnetic nanoparticles moving slightly faster than a single-wire system, and as a result a slightly faster magnetophoresis rate is observed. Interestingly, as the number of wires is increased to three, the evolution of concentration of the paramagnetic particles does not significantly differ from that of the two-wire configuration. In contrast, for diamagnetic nanoparticles, the rate of particle depletion within the domain increases as the number of wires increases to two and three. As shown in Fig.~S6 of the supplementary materials, diamagnetic nanoparticles are repelled from the flanks of the wires but can be attracted to their top or bottom surfaces. We hypothesize that the addition of more wires increases the available surface area at the tops of the wires, enhancing the attraction of diamagnetic particles to these regions. This effect likely contributes to the observed increase in the rate of diamagnetic nanoparticle depletion.}\par

\begin{figure*}
 \centering
 \includegraphics[trim=0cm 4cm 0cm 0cm,clip,width=16cm]{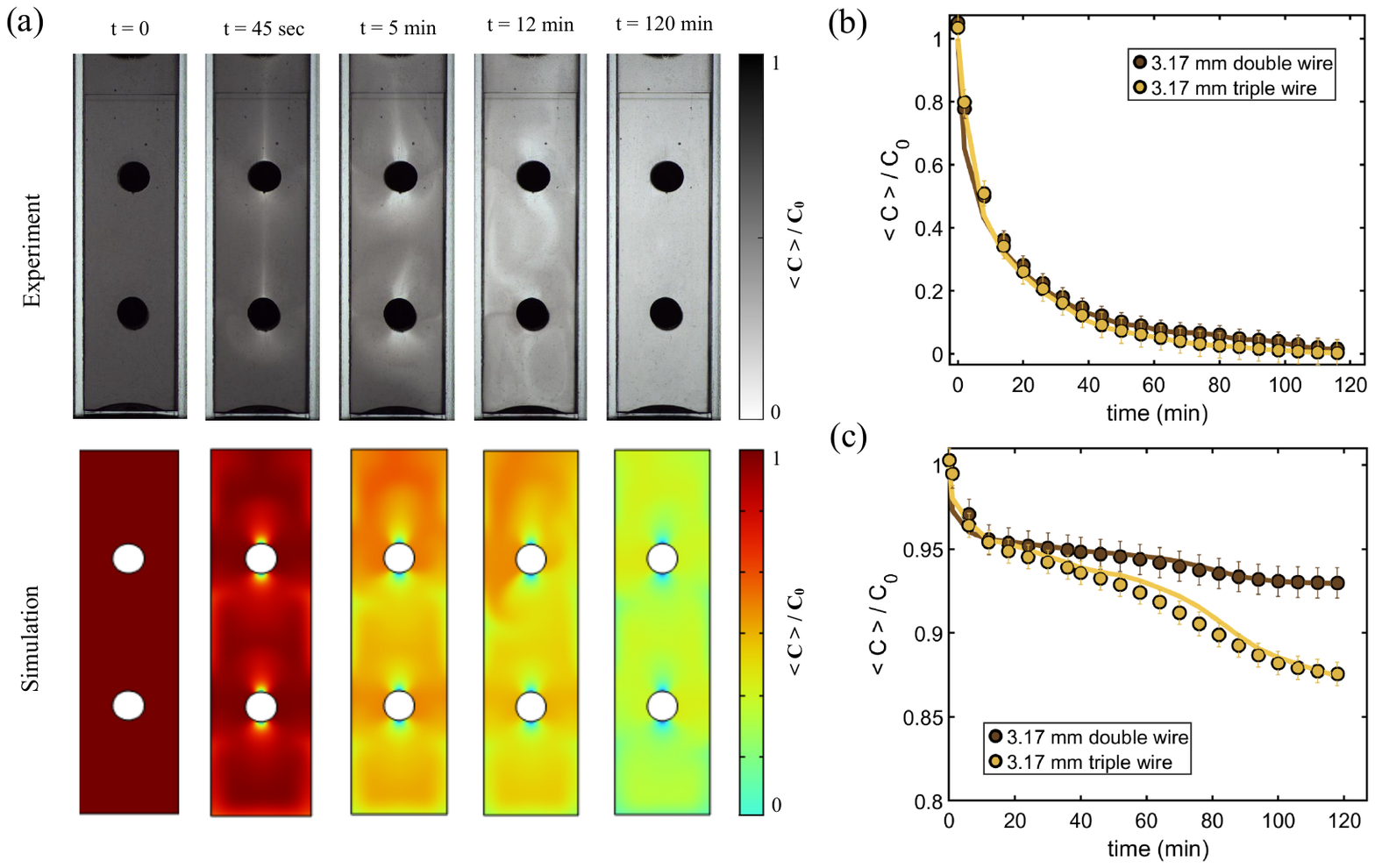}
 \caption{(a) The spatio-temporal evolution of particle concentration around the wires in experiments (top row), and corresponding simulations that best match the experiments (bottom row) for manganese oxide particles at a fixed initial concentration of 100 mg/L, a magnetic field of 1 T, and the wire diameter of d = 3.17 mm. The temporal evolution of averaged concentration for manganese oxide (b), and bismuth oxide nano-particles (c) as a function of the number of wires. }
 \label{fig:Fig9}
\end{figure*}
\hspace{0.5cm}
\HM{\subsection{Field-induced particle clustering}}
{Thus far, we have investigated the magnetophoresis of both paramagnetic and diamagnetic nano-particles in a closed cell, and our direct comparison between experiments and simulations suggests the formation of field-induced clusters under some conditions of initial concentration, magnetic field strength, magnetic susceptibility and wire diameter. Magnetic particles are well known to form clusters under strong magnetic fields~\cite{iacovita2020magnetic,kralj2015magnetic}}. Magnetic field-induced cluster formation is associated with the magnetic dipole interaction between the nanopartciles~\cite{chuan2012agglomeration,medvedeva2013sedimentation,vikesland2016aggregation,tsouris1995flocculation,faraudo2016predicting,faraudo2013understanding,andreu2011aggregation,liu1995field,de2008low}. The field-induced cluster formation can be understood in the context of two dimensionless parameters; aggregation parameter $N^*$ and coupling parameter $\Gamma$. The coupling parameter $\Gamma$ accounts for the balance between magnetic forces and thermal energy and is defined as\cite{tsouris1995flocculation,faraudo2016predicting,faraudo2013understanding,andreu2011aggregation,liu1995field,de2008low}: \begin{equation}
     \Gamma = \frac{\pi {\Delta\chi}^2 B^{2} {R_{p}}^3}{9 \mu_{o} k_{B} T}.
\end{equation}
In addition, the aggregation parameter $N^{*}$ is defined as~\cite{leong2020unified}: 
\begin{equation}
     N^{*} = \sqrt{\phi_{o} e^{\Gamma-1}}.
\end{equation}
Here, $\phi_{o}$ is the volume fraction of the nano-particles in the solution. The necessary condition for the formation of field-induced aggregation is $N^{*} >1$ and $\Gamma>1$.\par 

\HM{ Figure~\ref{fig:Surf_plot} presents the calculated values of the relevant dimensionless parameters as functions of particle size and magnetic field intensity. For paramagnetic manganese oxide particles at an initial concentration of $C_0 = 100$ mg/L, the analysis predicts that nanoparticle clustering should occur when the particle radius exceeds $R_p > 250$ nm and the applied magnetic flux density exceeds $\mathbf{B} > 0.6$ T. It is important to note that for smaller particle sizes that exists in the solutions of these experiments, the critical magnetic field required for the onset of cluster formation shifts to higher values. For example for $R_p < 175$ nm, the critical magnetic field for onset of cluster formation occurs when $\mathbf{B} > 1$ T.  Consequently, the larger particles present within the experimental size distribution are more susceptible to field-induced clustering.  However, both experimental observations and numerical simulations in the presence of wire (see Table~\ref{particlesize_B}) indicate that field-induced cluster formation begins at significantly lower magnetic field strengths, around $\mathbf{B} = 0.25$ T. This discrepancy highlights a limitation of the theoretical framework based on the dimensionless numbers $\Gamma$ and $N^*$, which assume idealized conditions such as spatially uniform magnetic fields. In practice, several additional factors, including strong magnetic field gradients, Derjaguin-Landau-Verwey-Overbeek (DLVO)-type interparticle interactions, and particle-fluid coupling, may influence clustering behavior and contribute to the earlier onset of field-induced aggregation observed in experiments involving a wire and an applied magnetic field.}\par 

\begin{figure} [H]
 \centering
 \includegraphics[width=1\linewidth]{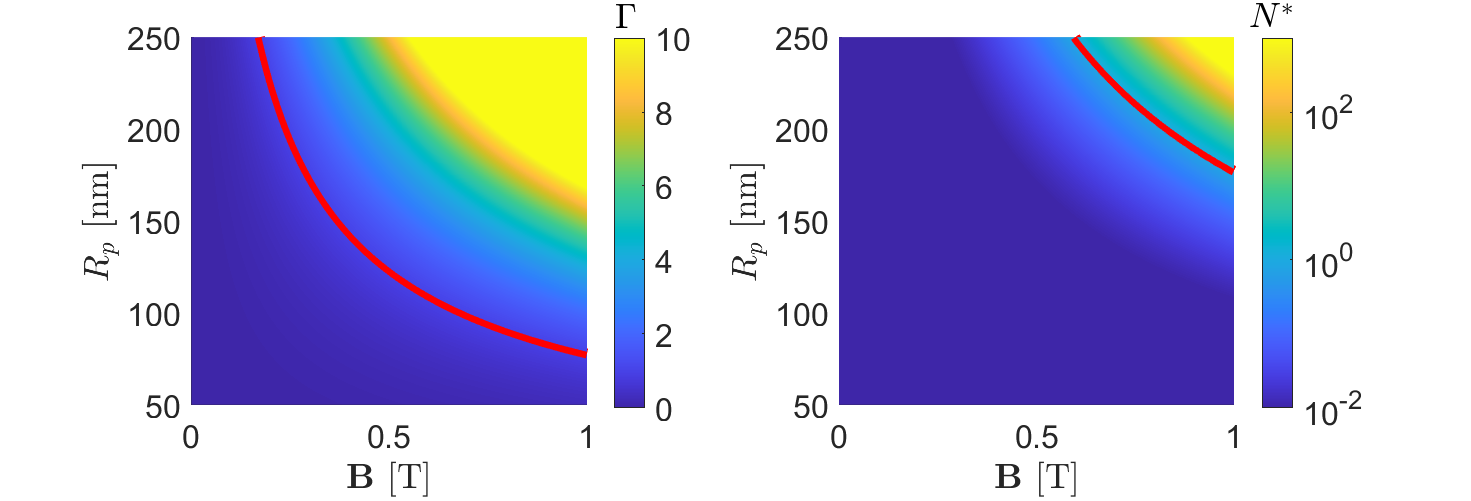}
 \caption{Surface plots of non-dimensional coupling parameter $\Gamma$ (left panel) and aggregation parameter $N^{*}$ (right panel) as a function of particle size $R_{p}$ and magnetic field $\mathbf{B}$ for MnO$_2$ particles with $C_0$ = 100 mg/L. The red isosurfaces indicate values of $N^{*}$ = 1 and $\Gamma$ = 1.}
 \label{fig:Surf_plot}
\end{figure}

\HM{To further probe the role of magnetic field non-uniformity in cluster formation, we conducted complementary experiments using a cuvette without the wire, but subjected to a uniform magnetic field of 1 T at the initial particle concentration of $C_0 = 100$ mg/L. As shown in Fig.~S7(a) of the supplementary materials, these experiments revealed the development of limited re-circulating flow zones within the cuvette. Particle depletion in this case was faster than in the zero-field condition ($\mathbf{B} = 0$ T), but still substantially slower than in experiments with the wire present (see Fig.~S7(b)). Our simulations that are best matched with experiments suggest that this intermediate behavior in the absence of wire arises from field-induced particle clustering (see Table~SI in the supplementary materials). This interpretation aligns well with the trends predicted in Fig.~\ref{fig:Surf_plot}, which indicates that cluster formation is expected for particles with radii greater than $R_p > 250$ nm when the applied uniform magnetic flux density exceeds $\mathbf{B} > 0.6$ T.} In our study, the magnetic field gradients around individual wires are remarkably high. In the presence of such gradients, a substantial magnetophoretic force emerges that may actively drive the nanoparticles toward the wire and enhance their mutual attraction. \HM{In addition, this strong magnetophoretic force induces flows on the surrounding fluid and the induced convection could significantly impact the field-induced cluster formation}. These results underscore the critical role of magnetic field gradients} \HM{and induced flows} in controlling nano-particle transport and aggregation.\par 

\HM{In deriving the above dimensionless criteria for field-induced cluster formation, DLVO-type interactions, such as van der Waals attraction and electrostatic double-layer repulsion, were not explicitly considered. However, our experimental observations under no field indicate that the particle sizes are larger than those reported by the vendor, suggesting a greater likelihood of interparticle interactions. Therefore, the DLVO forces may also play a non-negligible role in promoting field-induced clustering, particularly by enhancing the propensity for aggregation under applied magnetic fields.}\par 

\HM{Finally, we return to the results concerning diamagnetic bismuth oxide nanoparticles. The slight field-induced clustering observed in the presence of the wire at an applied magnetic field of $\mathbf{B} = 1$ T (as reported in Table~\ref{particlesize_B}) is not predicted by the current dimensionless framework, which was primarily developed for paramagnetic systems. To evaluate the role of magnetic field non-uniformity in this behavior, we conducted additional experiments at $\mathbf{B} = 1$ T in the absence of the wire (see Fig.~S7(c,d) in the supplementary materials). These results showed a much smaller degree of particle depletion compared to the wire-present case, but still more than the zero-field control ($\mathbf{B} = 0$ T). Our simulations under uniform field conditions (with no wire present) suggest a low likelihood of field-induced clustering in bismuth oxide. This slight tendency toward aggregation could be attributed to non-negligible DLVO-type interactions between bismuth oxide nano-particles. However, in the presence of the wire and consequently, a strong magnetic field gradient, the magnetophoretic force may significantly enhance the collision frequency between particles. Once brought into close proximity by these gradient-induced magnetophoretic forces, DLVO interactions can facilitate weak but observable cluster formation even in a diamagnetic system such as bismuth oxide.} To the best of our knowledge, a theoretical framework for field-induced cluster formation under a non-uniform magnetic field is not available. Hence, future studies on field-induced cluster formation should refine particle interaction models to explicitly incorporate magnetophoretic effects in non-uniform magnetic fields \HM{and also take into account the role of induced flows on particle-particle aggregation dynamics}. Doing so will enable a more comprehensive understanding of clustering mechanisms, particularly in systems with strong magnetic field gradients~\cite{tsouris1995flocculation}.\par

\section{\label{Con} Summary and Conclusions}
{In summary, we investigated the magnetophoresis behavior of paramagnetic manganese oxide and diamagnetic bismuth oxide nano-particle suspensions around a magnetized wire via a combination of experiments and multiphysics simulations. The key findings of this study can be summarized as follows:\par 

Our results demonstrate that paramagnetic nanoparticles exhibit a pronounced magnetophoresis behavior, becoming strongly attracted to the peripheries of the wire, whereas diamagnetic nanoparticles are repelled and tend to accumulate at the top of the wire. For paramagnetic nanoparticles, the strong magnetic force induces secondary flows throughout the domain, resulting in vortex formation that further enhances nanoparticle capture by the wire. As the magnetic field strength increases, the magnetophoretic effect becomes more pronounced, resulting in a maximum capture of 50\% for MnO$_2$ particles after 2 hours at 1 T, with a wire diameter of 0.8 mm. In contrast, diamagnetic nanoparticles exhibit moderate concentration variations ($\approx$ 5\%) up to 0.75 T, followed by a sharp increase in 20\% capture at 1 T. The depletion rate of paramagnetic nanoparticles within the domain is proportional to the wire diameter, whereas diamagnetic nanoparticles are captured at a slower rate as the wire diameter increases. Introducing a second or third wire slightly improves the capture efficiency for both manganese oxide and bismuth oxide nanoparticles, with an observed enhancement of approximately 10\%.\par  

Furthermore, multiphysics simulations suggest that paramagnetic nanoparticles may undergo field-induced cluster formation beyond critical thresholds of magnetic field strength ($B \geq$0.25 T) and nanoparticle concentrations. This clustering accelerates the magnetophoresis rate beyond what would be expected from equilibrium particle size distributions. Interestingly, at the highest imposed magnetic field, diamagnetic nanoparticles may also experience moderate cluster formation \HM{due to non-negligible DLVO-type forces}. To the best of our knowledge, this phenomenon has not been previously reported in the literature. Finally, this study reveals that strong magnetic field gradients generate significant magnetophoretic forces that may drive nanoparticles toward the wire\HM{, induce flows} and facilitate cluster formation at lower field strengths than predicted by uniform-field models based on $N^*$ and $\Gamma$ parameters. These findings highlight the need to refine interaction models to account for magnetophoretic effects in non-uniform magnetic fields.}

\HM{\section{Supplementary Materials}
Supplementary materials show additional experimental and simulation data on magnetophoresis of particles under influence of the magnetic field.}

\section{Acknowledgments}

This work was performed at the National High Magnetic Field Laboratory, which is supported by the National Science Foundation Cooperative Agreement No. DMR-1644779 and the state of Florida. This work was supported by the Center for Rare Earths, Critical
Minerals, and Industrial Byproducts, through funding provided by the State of Florida. \HM{We would like thank Dana Ezzeddine, Daniel Barzycki, and the Ricarte Lab for assistance with dynamic light scattering measurements and use of their instrument.}

 \bibliographystyle{elsarticle-num} 
 \bibliography{cas-refs}





\end{document}


\singlespacing

\title{Supplementary Materials: Magnetophoresis of weakly magnetic nano-particle suspension around a wire}
\author{Mohd Bilal Khan}
\affiliation{Department of Chemical and Biomedical Engineering, FAMU-FSU College of Engineering, Tallahassee, FL, 32310, USA}
\affiliation{Center for Rare Earths, Critical Minerals, and Industrial Byproducts, National High Magnetic Field Laboratory, Tallahassee, FL 32310, USA}

\author{Peter Rassolov}
\affiliation{Department of Chemical and Biomedical Engineering, FAMU-FSU College of Engineering, Tallahassee, FL, 32310, USA}
\affiliation{Center for Rare Earths, Critical Minerals, and Industrial Byproducts, National High Magnetic Field Laboratory, Tallahassee, FL 32310, USA}
\author{Jamel Ali}
\affiliation{Department of Chemical and Biomedical Engineering, FAMU-FSU College of Engineering, Tallahassee, FL, 32310, USA}
\affiliation{Center for Rare Earths, Critical Minerals, and Industrial Byproducts, National High Magnetic Field Laboratory, Tallahassee, FL 32310, USA}
\author{Theo Siegrist}
\affiliation{Department of Chemical and Biomedical Engineering, FAMU-FSU College of Engineering, Tallahassee, FL, 32310, USA}
\affiliation{Center for Rare Earths, Critical Minerals, and Industrial Byproducts, National High Magnetic Field Laboratory, Tallahassee, FL 32310, USA}
\author{Munir Humayun}
\affiliation{Center for Rare Earths, Critical Minerals, and Industrial Byproducts, National High Magnetic Field Laboratory, Tallahassee, FL 32310, USA}
\affiliation{Department of Earth, Ocean and Atmospheric Science, Florida State University, Tallahassee, FL 32304, USA.}

\author{Hadi Mohammadigoushki}
\thanks{Corresponding author}\email{hadi.moham@eng.famu.fsu.edu}
\affiliation{Department of Chemical and Biomedical Engineering, FAMU-FSU College of Engineering, Tallahassee, FL, 32310, USA}
\affiliation{Center for Rare Earths, Critical Minerals, and Industrial Byproducts, National High Magnetic Field Laboratory, Tallahassee, FL 32310, USA}
\date{\today}

\maketitle
\section{Calibration curve}
\begin{figure}[H]
 \centering
 \includegraphics[trim=0cm 8.5cm 0cm 0cm,clip,width=15cm]{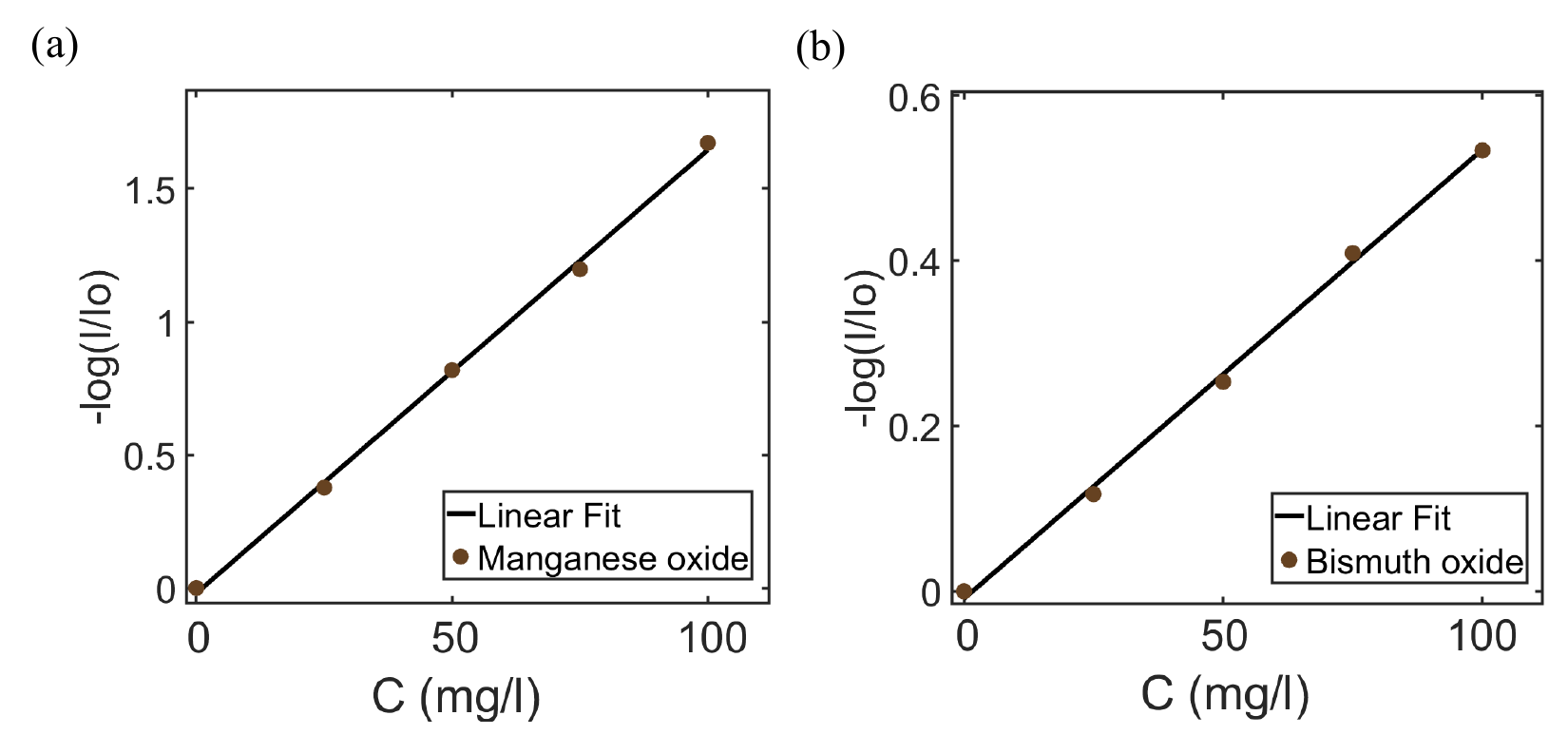}
 \caption{Normalized absorbance intensity as a function of initial nano-particle concentration for (a) manganese oxide and (b) bismuth oxide under identical experimental conditions. These graphs serve as calibration curves that are used in further analysis of the spatio-temporal evolution of particle concentration under influence of the external magnetic field.}
 \label{fig:Calib}
\end{figure}

To assess the concentration variation within the experimental domain, we measured the absorbance intensity (light intensity) at a given concentration. Fig~\ref{fig:Calib} shows the normalized absorbance intensity as a function of concentration measured for both nano-particles. In here $I$ is the intensity of the sample that has a solution and $I_{o}$ is the background intensity (without sample). Clearly, the normalized light intensity shows a linear dependency with respect to the initial particle concentration, which is consistent with the Beer-Lambert law. 
 \begin{equation}
     -log (I/I_{o}) = A 
 \end{equation}
The linear fit regression has been performed by the relationship between absorbance $(A)$ and concentration $(C)$, which states~\cite{mayerhofer2016employing}:
 \begin{equation}
     A = \epsilon C l 
 \end{equation}
 where $\epsilon$ is the molar absorptivity and $l$ is the path length of the sample. For linear fit, equation 2 can be expressed as:
 \begin{equation}
     A = m C + b 
 \end{equation}
  Here, the slope of the linear fit $m$ provides the product of $\epsilon \cdot l$, which is used to calculate the absorptivity, while $b$ is the intercept. 

\section{Analytical form of magnetic field distribution around a single wire}
The magnetic flux density ($\mathbf{B}$) and its gradient ($\mathbf{B} \cdot \nabla)\mathbf{B}$ around the wire was calculated analytically using the two equations derived before\cite{svoboda2004magnetic}. These equations define the components of the magnetic field around the wire as:~\cite{svoboda2004magnetic}
\begin{equation}
    H_{r} = H_{o} (1+K \frac{a^2}{r^2}) cos \theta 
\end{equation}
\begin{equation}
    H_{\theta} = H_{o} (1-K \frac{a^2}{r^2}) sin \theta 
\end{equation}
Where $H_{r}$ and $H_{\theta}$ are the radial and angular components of the magnetic field strength, $a$ is the wire diameter and, $r$ is the distance of the particle from the wire. If $H_{o} < H_{s}$, $K =1$, while for $H_{o} > H_{s}$, $K=\frac{M_{s}}{2 H_{o}}$. Here, $M_{s}$ and $H_{s}$ are the saturation magnetization and saturation magnetic field strength of the wire material, respectively. 
\begin{figure}[H]
 \centering
 \includegraphics[trim=0cm 0cm 8cm 0cm,clip,width=12cm]{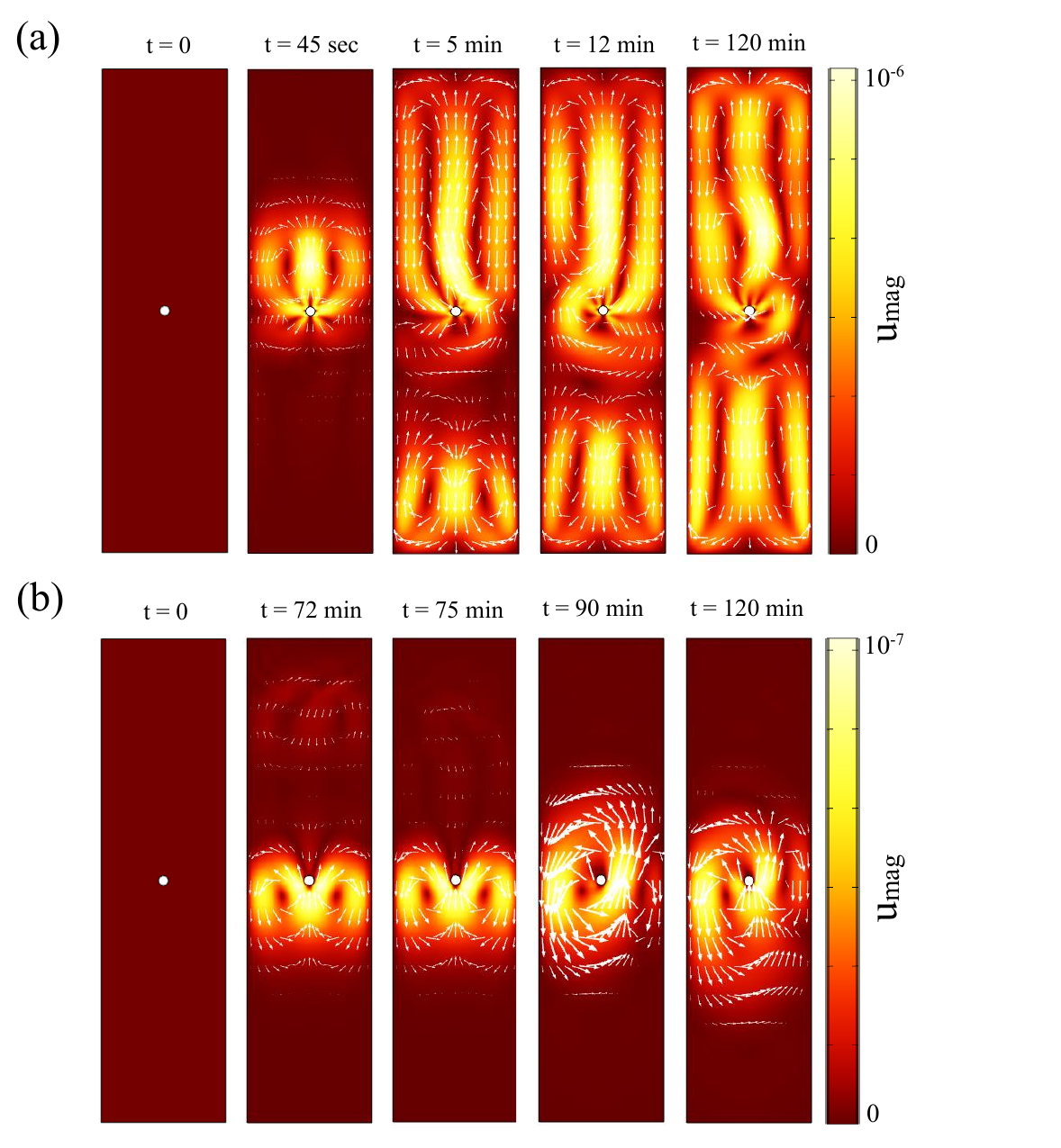}
 \caption{2D velocity vectors for (a) manganese oxide and (b) bismuth oxide at a concentration = 100 mg/L, magnetic field $\mathbf{B}$ = 1 T, and a wire diameter, $d$ = 0.88 mm.}
 \label{fig:vel_mag}
\end{figure}

\begin{figure} [H] 
 \centering
 \includegraphics[trim=0cm 0cm 0cm 0cm,clip,width=17cm]{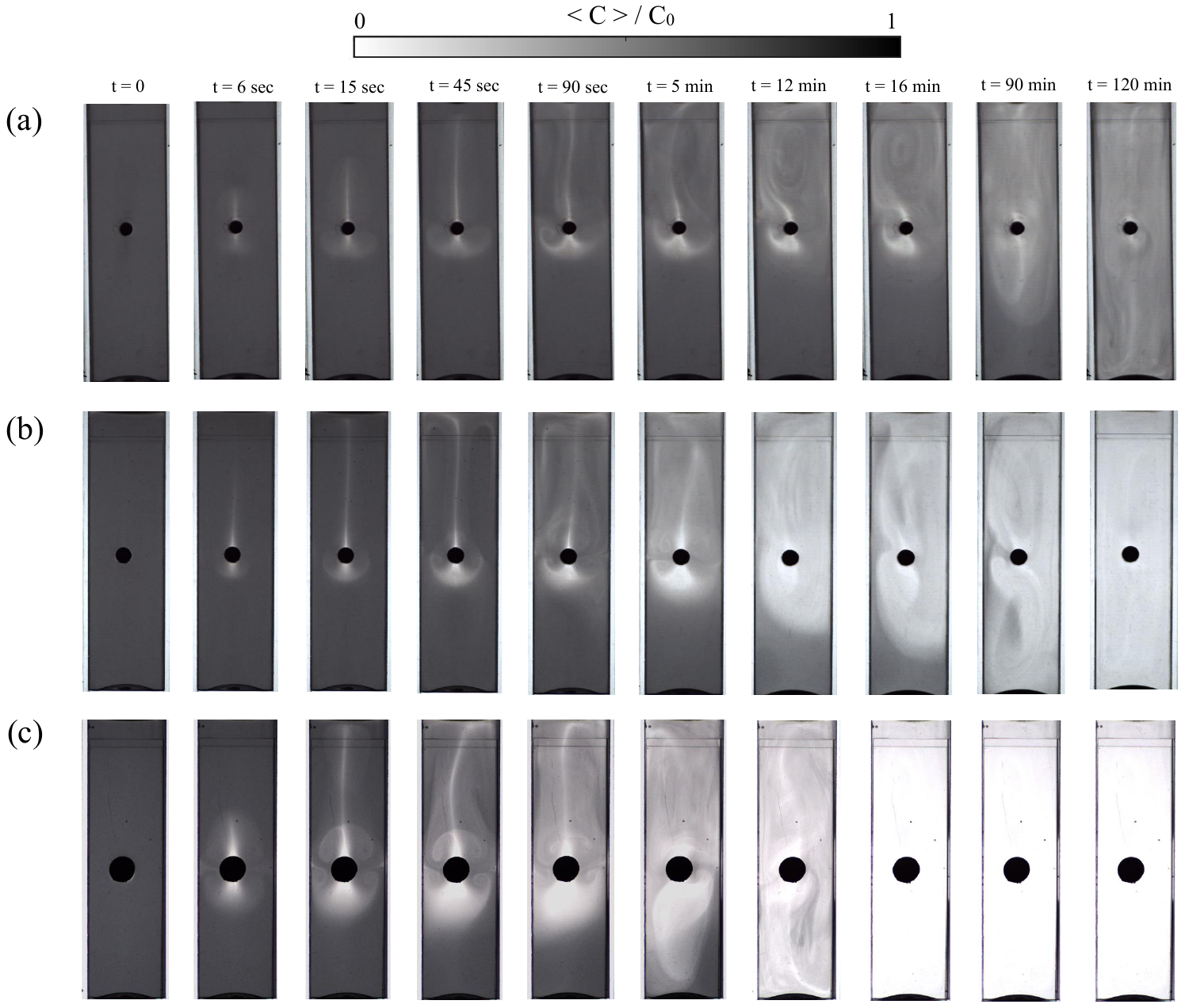}
 \caption{Spatio-temporal evolution of particle concentrations around a wires with (a) $d$ = 1.62 mm, (b) $d$ = 2 mm, and (c) $d$ = 3.17 mm for manganese oxide nanoparticles  at a fixed concentration of 100 mg/L, and a magnetic field of $\mathbf{B}$ = 1 T.}
 \label{fig:Fig5}
\end{figure}

\begin{figure} [H] 
 \centering
 \includegraphics[trim=0cm 0cm 0cm 0cm,clip,width=17cm]{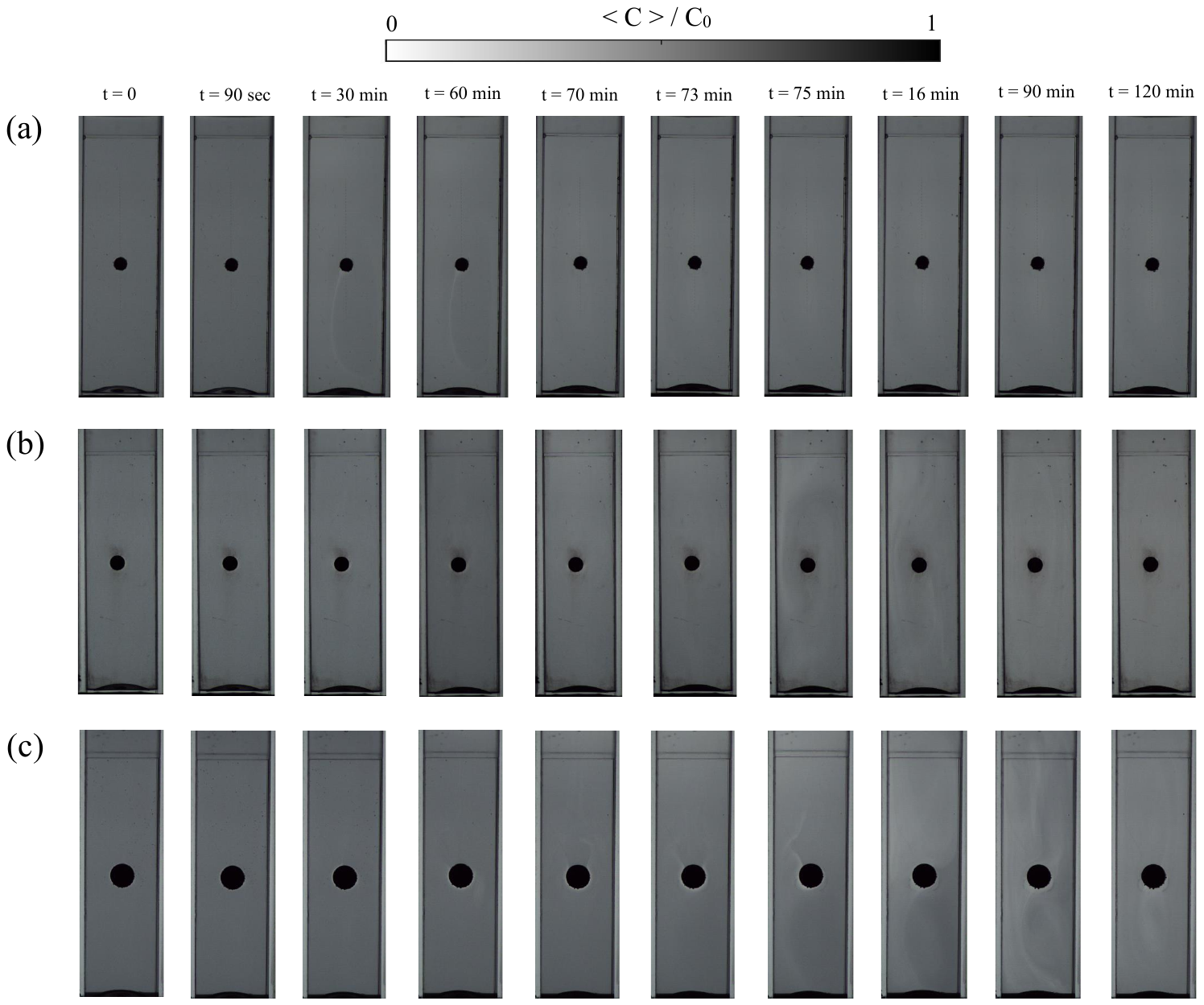}
 \caption{Spatio-temporal evolution of particle concentrations around a wires with (a) $d$ = 1.62 mm, (b) $d$ = 2 mm, and (c) $d$ = 3.17 mm for bismuth oxide nanoparticles  at a fixed concentration of 100 mg/L, and a magnetic field of $\mathbf{B}$ = 1 T.}
 \label{fig:Fig5}
\end{figure}

\begin{figure} [H]
 \centering
 \includegraphics[trim=0cm 6cm 0cm 0cm,clip,width=17cm]{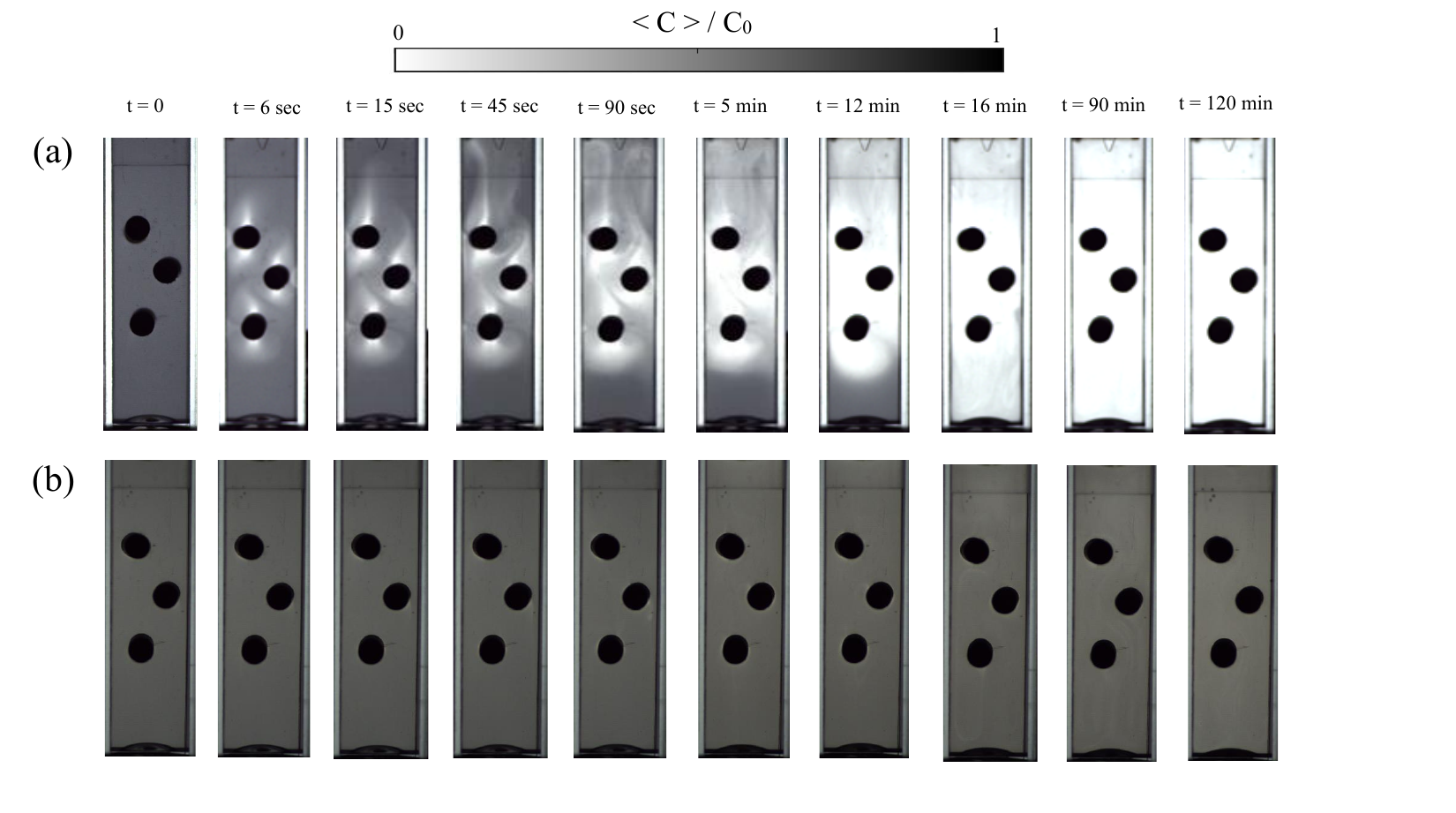}
 \caption{Effect of multiple wires on magnetic separation of (a) manganese oxide and (b) bismuth oxide nanoparticles at a fixed concentration of 100 mg/L, magnetic field of $\mathbf{B}$ = 1 T and a wire diameter $d$ = 3.17 mm.}
\end{figure}  

\begin{figure}[H]
 \centering
 \includegraphics[trim=0cm 6cm 0cm 0cm,clip,width=17cm]{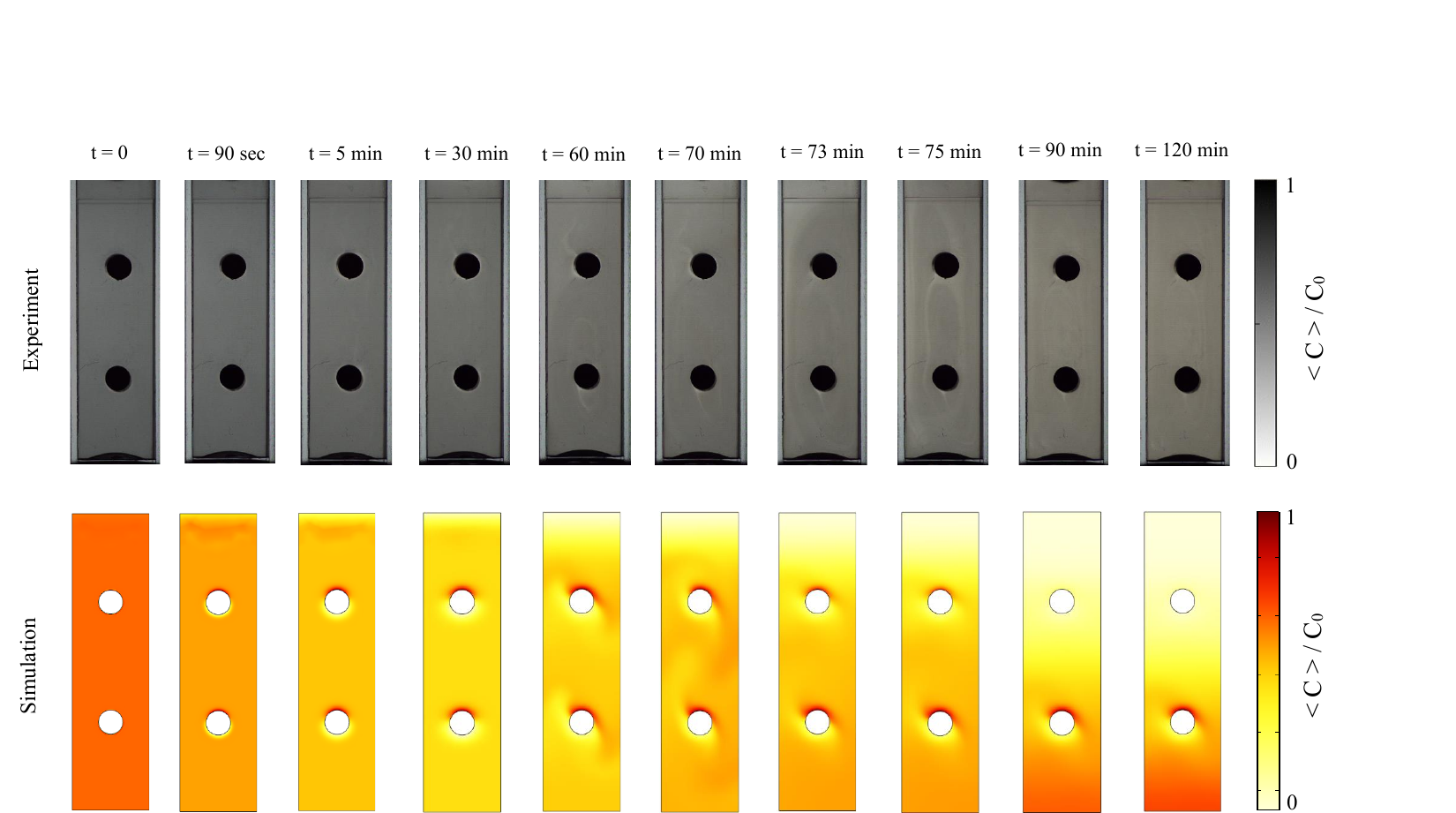}
 \caption{Spatio-temporal evolution of bismuth oxide particles at a fixed concentration of 100 mg/l, magnetic field of $\mathbf{B}$ = 1 T and a wire diameter $d$ = 3.17 mm for (a) experiments and (b) corresponding numerical simulations.}
 \label{fig:Fig6}
\end{figure}

\begin{figure} [H]
 \centering
 \includegraphics[trim=0cm 0cm 6cm 0cm,clip,width=14cm]{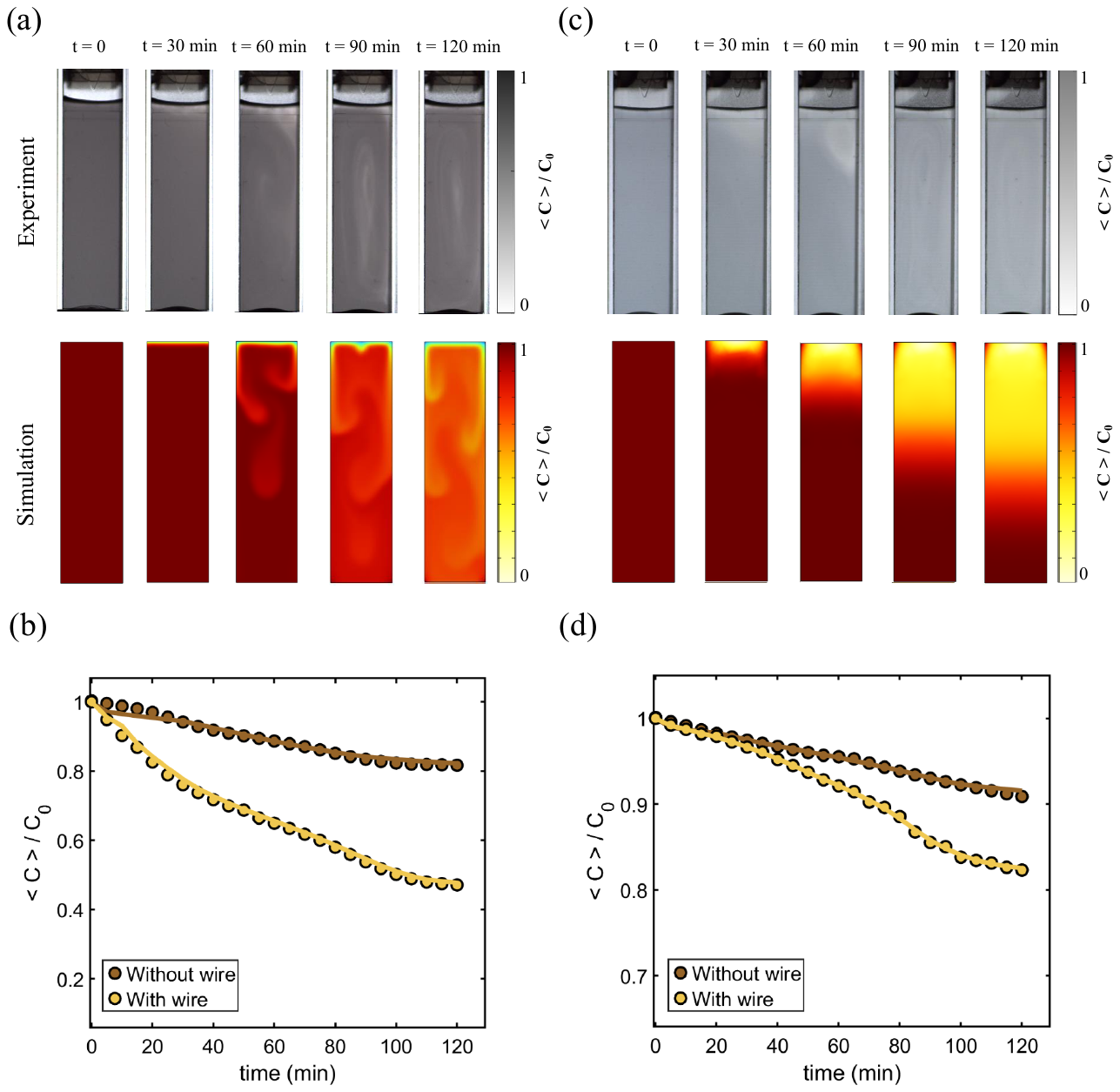}
 \caption{\textcolor{blue}{Spatio-temporal evolution of particle concentrations measured in experiments (top row), and calculated (bottom row) for manganese oxide (a) and bismuth oxide particles (c) at a concentration = 100 mg/L, and a uniform magnetic field $\mathbf{B}$ = 1 T under no gradient condition. (b) and (d) Volume-averaged normalized concentration as a function of time for experiments (symbols), and numerical simulations (lines). It is noted that in experiments with bismuth oxide particles, some flow patterns emerge near the top of the cuvette. Simulations predict the formation of a concentration gradient along the cuvette's longest axis; however, this gradient is not clearly observable in the experimental data. }}
 \label{fig:Surf_plot}
\end{figure}

\begin{table}[h!]
\centering
\begin{tabular}{|c|c|c|c|c|}
\hline
\textbf{C$_{0}$} & \multicolumn{2}{c|}{\textbf{Manganese oxide}} & \multicolumn{2}{c|}{\textbf{Bismuth oxide}} \\
\cline{2-5}
\textbf{(mg/l)} & $\phi_i$ & $R_{pi}$ [nm] & $\phi_i$ & $R_{pi}$ [nm] \\
\hline
100 & 0.72 0.18 0.10 & 150 400 600 & 0.73 0.15 0.12 & 80 120 160 \\
\hline
\end{tabular}
\caption{\textcolor{blue}{Particle size distribution for simulations performed for no wire condition.}}
\label{tab:phi_Rp_data}
\end{table}

 \bibliography{cas-refs} 

 \bibliographystyle{unsrt}